\documentclass[prd,twocolumn,nofootinbib,preprintnumbers,superscriptaddress,showpacs]{revtex4-1}

\usepackage{amsfonts}
\usepackage{amsmath}
\usepackage{amssymb}
\usepackage{bm}
\usepackage{dcolumn}
\usepackage{graphicx}
\usepackage{graphics}
\usepackage[latin1]{inputenc}
\usepackage{latexsym}
\usepackage{rotating}
\usepackage{url}
\usepackage{xspace}
\usepackage[usenames]{color}
\usepackage{mathrsfs}
\usepackage{hyperref}
\usepackage{epstopdf}
\usepackage{tensor}
\usepackage[margin=1.0in]{geometry}
\usepackage[scientific-notation=true]{siunitx}
\usepackage{placeins}

\newcommand{\be}{\begin{equation}}
\newcommand{\ee}{\end{equation}}
\newcommand{\bal}{\begin{align}}
\newcommand{\bsp}{\begin{split}}
\newcommand{\esp}{\end{split}}

\newcommand{\GR}{{\mbox{\tiny GR}}}
\newcommand{\Hz}{{\mbox{\tiny H}}}

\newcommand{\eff}{{\mbox{\tiny eff}}}
\newcommand{\nonlin}{{\mbox{\tiny nonlin}}}

\newcommand{\SH}{{\mbox{\tiny SCHW}}}

\newcommand{\SPH}{{\mbox{\tiny SPH}}}
\newcommand{\xT}{{\mbox{\tiny x}}}

\newcommand{\bw}{\begin{widetext}}
\newcommand{\ew}{\end{widetext}}

\newcommand{\pd}{\partial}
\newcommand{\cd}{\nabla}
\newcommand{\tn}{\tensor}
\newcommand{\tnst}{\tensor*}

\newcommand{\ssqth}{\sin^2{\theta}}

\newcommand{\mcl}{\mathcal}

\newcommand{\mrm}{\mathrm}
\newcommand{\rarr}{\rightarrow}

\newcommand{\lb}{\left(}
\newcommand{\rb}{\right)}

\newcommand{\lsb}{\left[}
\newcommand{\rsb}{\right]}

\begin{document}
\title{Numerical Black Hole Solutions in Modified Gravity Theories: \\ Axial Symmetry Case}

\author{Andrew Sullivan}
\affiliation{Department of Physics, Montana State University, Bozeman, MT 59717, USA}
\author{Nicol\'as Yunes}
\affiliation{Department of Physics, University of Illinois at Urbana-Champaign, Urbana, IL 61801, USA}
\author{Thomas P.~Sotiriou}
\affiliation{School of Mathematical Sciences \& School of Physics and Astronomy, University of Nottingham,
University Park, Nottingham, NG7 2RD, United Kingdom}

\date{\today}

\begin{abstract} 

We extend a recently developed numerical code to obtain stationary, axisymmetric solutions that describe rotating black hole spacetimes in a wide class of modified theories of gravity. The code utilizes a relaxed Newton-Raphson method to solve the full nonlinear modified Einstein's Equations on a two-dimensional grid with a Newton polynomial finite difference scheme. We validate this code by considering static and axisymmetric black holes in General Relativity. We obtain rotating black hole solutions in scalar-Gauss-Bonnet gravity with a linear (linear scalar-Gauss-Bonnet) and an exponential (Einstein-dilaton-Gauss-Bonnet) coupling and compare them to analytical and numerical perturbative solutions. From these numerical solutions, we construct a fitted analytical model and study observable properties calculated from the numerical results.

\end{abstract}

\maketitle

\section{Introduction}
\label{sec:intro}

As we enter a new era of multi-messenger astrophysics, many new experiments will allow us to test Einstein's theory of general relativity (GR) in the strong field regime~\cite{Yunes2013,Gair2013,Will2014,Psaltis2008,Damour2009,Barack:2018yly}. Strong field observations test GR by probing whether the properties of astrophysical compact objects match GR's prediction. However, one wants not only to test whether GR predictions fit the data, but also whether they do so better than potential alternatives. This requires the study of compact objects in modified gravity, and in particular, the solution to the full field equations for realistic astrophysical black holes. Although approximate solutions might provide a simplification to the complexity of the modified field equations, this simplification comes at the expense of accuracy. As the precision of our observations is improving, so should the precision of our modeling, which therefore motivates a fully numerical study. 

Such numerical solutions serve multiple purposes. On the one hand, they can be used to study the stability of black holes in modified gravity.  For example,  in Einstein-dilaton-Gauss-Bonnet gravity~\cite{CAMPBELL1992199, MIGNEMI1993299, PhysRevD.54.5049, Sotiriou:2013qea,Yunes2013, 0264-9381-32-24-243001,Barack:2018yly}, stationary spacetimes have been used to study the properties of perturbed black holes through their quasinormal mode spectrum~\cite{Blazquez-Salcedo:2016enn}. On the other hand, numerical solutions can also be used directly to determine how certain observables deviate from GR's predictions. For example, the location of the innermost-stable-circular-orbit and of the light ring can be calculated from these numerical spacetimes, and these locations can be inferred from observations of accretion disks around black holes~\cite{Abramowicz2013} and from black hole shadows~\cite{0264-9381-35-23-235002} respectively. 

Several methods exist to numerically find black hole solutions in modified gravity, and we recently developed one such method that is applicable to a wide class of theories but only to static and spherically symmetric black holes~\cite{PhysRevD.101.044024}. Our infrastructure uses symbolic manipulation software to calculate the modified field equations and export them into an executable algorithm written in the C programming language. These equations are then discretized using a finite element method by replacing each differential operator at each grid point with a Newton interpolation polynomial and calculating the residual of the field equations. By minimizing this residual using a relaxed Newton-Raphson method, we can iteratively converge to the desired solution by calculating the linearized correction to our functions through the solution to a linear system of equations evaluated from the Jacobian matrix of our discretized differential equations. 

In this paper, we extend our numerical infrastructure to rotating (i.e.~stationary, axisymmetric and vacuum) black hole spacetimes.
We first validate our numerics by studying rotating black holes in GR and we directly compare the numerical result to the known Kerr solution. 
After this validation, we construct stationary, axially symmetric black holes in scalar Gauss-Bonnet (sGB) gravity, a well-motivated modified theory ~\cite{CAMPBELL1992199, MIGNEMI1993299, PhysRevD.54.5049, Sotiriou:2013qea} that is a member of the quadratic gravity class~\cite{Yunes2013, 0264-9381-32-24-243001,Barack:2018yly}. In the action of sGB gravity, a scalar couples to the Gauss-Bonnet invariant, $\mcl{G}$ through a coupling function $F(\psi)$. Different coupling functions have been considered in the literature: the exponential case is commonly referred to as Einstein-dilaton-Gauss-Bonnet (EdGB) gravity, while $F(\psi)=\psi$ is commonly referred to as the linear sGB gravity. 

Part of the motivation for considering sGB as the first example to study with this new code is that black holes in this class of theories have already received a lot of attention. Stationary black holes have been found in sGB assuming spherical symmetry \cite{CAMPBELL1992199, MIGNEMI1993299, PhysRevD.54.5049,Sotiriou:2013qea,PhysRevD.90.124063} or working perturbatively in slow-rotation~\cite{PhysRevD.90.044066, PhysRevD.84.087501, PhysRevD.92.083014}. Stationary, axisymmetric black holes with arbitraty spin in EdGB have only been obtained numerically~\cite{PhysRevD.93.044047}. There has also been recent work on the dynamical evolution of black holes and binaries in sGB gravity~\cite{PhysRevD.94.121503,Benkel:2016rlz,Witek:2018dmd,Ripley:2019hxt,Okounkova:2020rqw}. Recently, it has also been shown that when $F(\psi)$ is quadratic in $\psi$ certain models exhibit {\em black hole scalarization}: the black hole acquire scalar hair only when their mass or spin exceeds a certain threshold \cite{PhysRevLett.120.131104, PhysRevLett.120.131103,Dima:2020yac}.

We will construct fully nonlinear solutions in linear sGB and EDGB gravity that describe stationary and axially symmetric black holes. We will compare these solutions to perturbative ones found in a weak-coupling expansion $\bar{\alpha} = \alpha/\rho_{\Hz}^2 \ll 1$, where $\rho_{\Hz}$ the horizon radius. This allows us to verify that our numerical solutions in the non-rotating limit are equal to the analytically-known, spherically symmetric, perturbed solution, and to compare the fully nonlinear solutions to the perturbed weak-coupling expansion. We will also use these solutions to construct analytic, closed-form functions that are excellent approximations to our numerical solutions. We will conclude with an analysis of the properties of some physical observables that can be calculated with our non-linear solutions and our analytic, closed-form approximations. 


%
\vspace{-0.4cm}
\subsection*{Executive Summary}
\vspace{-0.2cm}
\begin{figure}[htb]
\begin{center}
\includegraphics[width=7.5cm,clip=true]{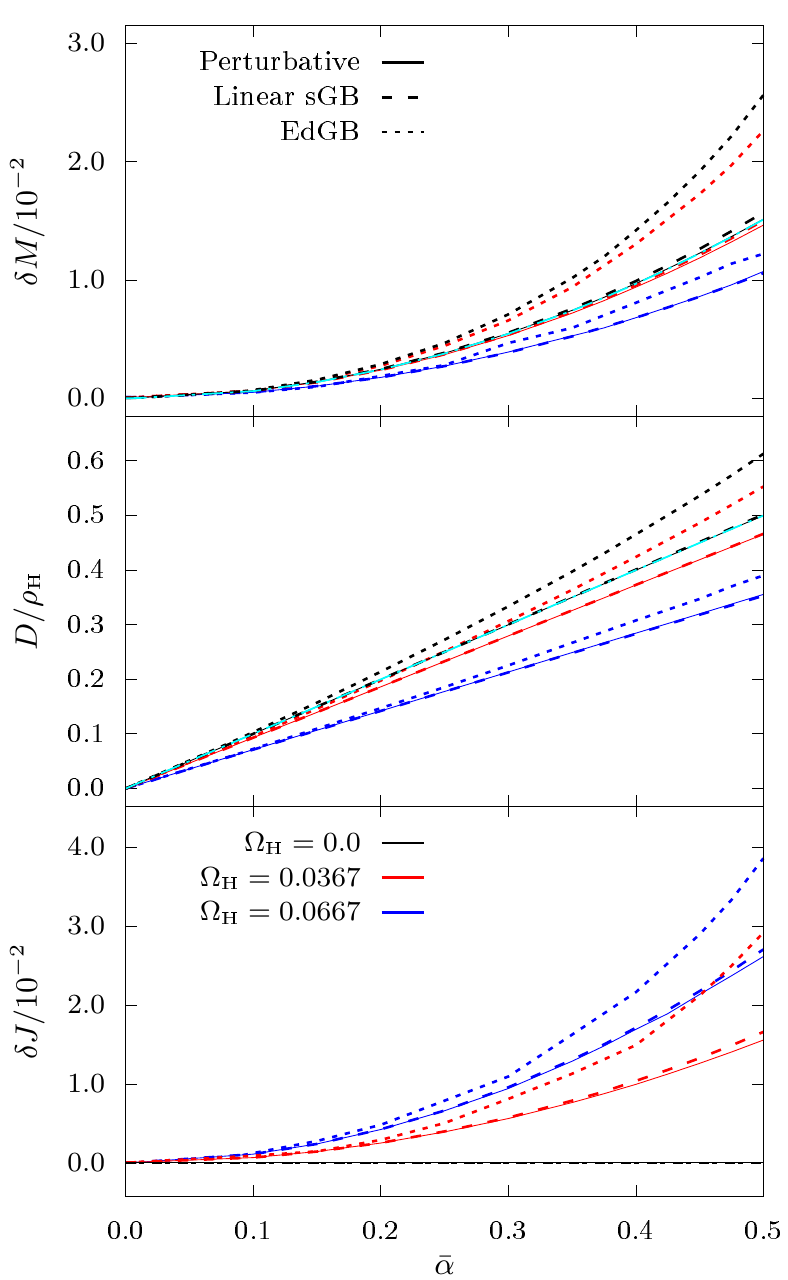}
\caption{\label{fig:MassChg}
(Color online)
Fractional change in ADM mass (top) and angular moment (bottom), and the dimensionless scalar charge (middle) versus dimensionless coupling $\bar{\alpha}$ for three event horizon angular velocities. Weak-coupling solutions are denoted with solid lines, linear sGB solutions with dashed lines, and EdGB solutions with dotted lines, while the analytic perturbative solution in spherical symmetry is shown with dashed lines. The weak-coupling solution agrees with the linear sGB solution, but disagrees with the EdGB solution. As we increase the spin of the black hole, the scalar charge decreases, and  correspondingly the GR deviations in the mass decrease, while the angular momentum increases.}
\end{center}
\end{figure}

One of the main products of our analysis is the extension of our numerical infrastructure from spherical~\cite{PhysRevD.101.044024} to axial symmetry, which we hereby openly release to the community.  This extension requires the discretization of our partial differential equations on a two dimensional grid and the replacement of each differential operator in the new dimension with a similar Newton interpolation polynomial. To discretize any mixed partial derivatives, we follow the approach of~\cite{SCHONAUER1989279} and introduce an auxiliary variable with a corresponding differential equation whose residual must simultaneously be minimized with the remaining system of equations. Validation using GR shows convergence to the Kerr solution with a tolerance of $10^{-5}$ in 4 iterations.  

With the code validated, we then move to a study of rotating black holes in linear sGB and EDGB gravity. In the non-rotating case, we recover the previous observation that the perturbative solution that assumes weak-coupling agrees exceptionally well with the exact solution in linear sGB, while there are still large differences with the solution in EdGB. When we include rotation, we find that the magnitude of these differences in the exponential coupling solution is suppressed as we increase the rotation of the black hole. This seemingly counter-intuitive behavior can be explained through Fig.~\ref{fig:MassChg}, which shows the relative fractional correction in the ADM mass (top) and angular momentum (bottom), and the scalar monopole charge (middle) as a function of the dimensionless sGB coupling parameter $\bar{\alpha}$. For the same $\bar{\alpha}$, increasing the angular velocity of the black hole horizon \textit{decreases} its scalar charge and \textit{suppresses} the deviation in the mass from its GR value, while \emph{increasing} the angular momentum.

\begin{figure}
\begin{center}
\includegraphics[width=7.5cm,clip=true]{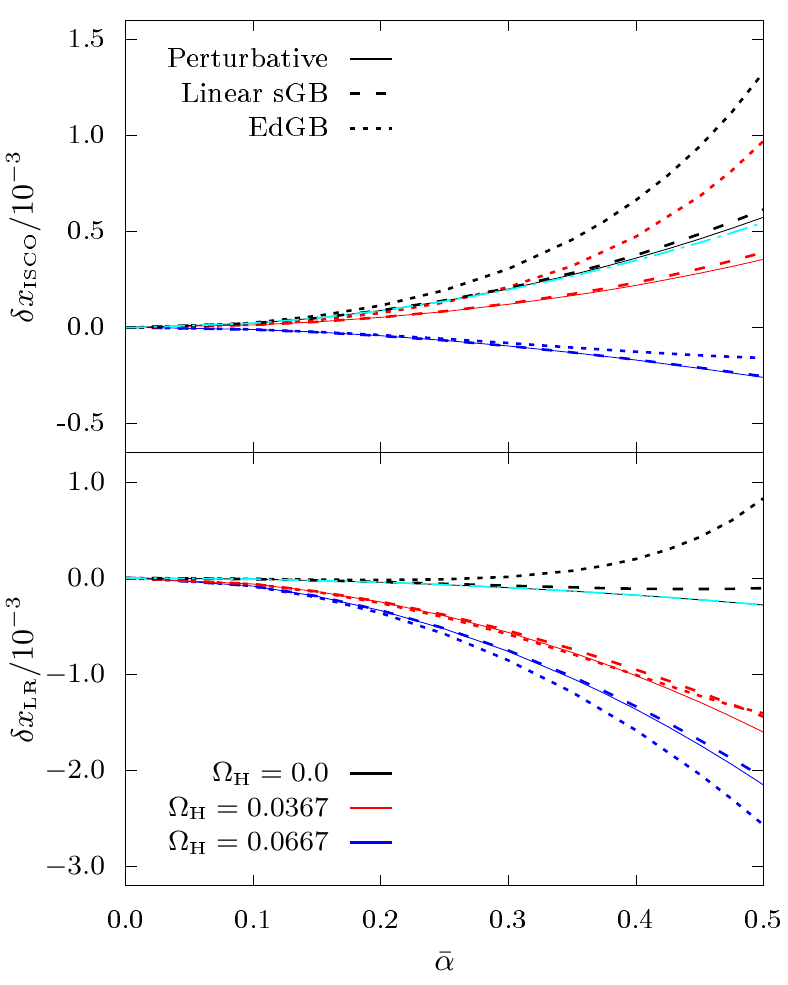}
\caption{\label{fig:ISCOLR}
(Color online)
Scaled fractional change in the location of the innermost stable circular orbit (top) and light ring (left) versus dimensionless coupling $\bar{\alpha}$ for three event horizon angular velocities, using the same conventions as in Fig.~\ref{fig:MassChg}. As before, observe that the observables computed with the weak-coupling solution agree with those of loinear sGB theory, while they disagree with the predictions of EdGB gravity for large $\bar{\alpha}$. The deviations from GR are suppressed as the spin increases, with the locations of the innermost stable circular orbit and light ring shifting closer to the horizon.}
\end{center}
\end{figure}

From the solutions, we calculate the location of the innermost stable circular orbit (ISCO) and the light ring, as shown in Fig.~\ref{fig:ISCOLR}. These observables, when computed with the weak-coupling solution, agree with those computed with the linear sGB solution, while they disagree with those calculated with the EdGB solution as $\bar{\alpha}$ increases. Observe also that as the rotation of the black hole increases, the location of the ISCO and the light ring decreases such that in the extremal limit it is coincident with the event horizon, as one also finds in GR. However, in linear sGB gravity an increase in $\bar{\alpha}$ increases both the mass and scalar charge of the black hole, which shifts the location of the ISCO and light ring to larger radii. We observe these two competing effects in Fig.~\ref{fig:ISCOLR}, where the rotation of the black hole reduces and eventually changes the sign of the fractional change in the location of both the ISCO and the light ring. 

%

Finally, we use the numerical solutions found in linear sGB and EdGB gravity to construct analytic, closed-form expressions for the four-dimensional spacetime metric that is capable of reproducing the numerical results to the accuracy of the latter. We provide the fitting coefficients for this analytic representation online, together with the Mathematica routine that provides the metric components themselves. This analytic representation now enables future studies of the stability of such black hole solutions. 

The numerical infrastructure is freely available to the scientific community to use as a tool to explore black hole spacetimes beyond GR. The generality of the numerics stems from the use of minimal assumptions about the specific modified theory of gravity considered, about the boundary conditions on the horizon and infinity, and about the existence of additional fields beyond the metric tensor. The analytical, closed-form representation can be used directly to calculate astrophysical observables, such as those associated with accretion disks around black holes~\cite{Abramowicz2013}, shadows~\cite{0264-9381-35-23-235002}, or quasinormal modes of black hole mergers~\cite{Blazquez-Salcedo:2016enn}. These analytic metrics can also serve as the basis for the construction of initial data for full numerical simulations of merging black holes in EDGB and sGB gravity.

\vspace{1cm} 

The remainder of this paper is organized as follows. 
Section~\ref{sec:NM} outlines adjustments that are required to extend the numerical algorithm to partial differential equations in axial symmetry.
Section~\ref{sec:GR} validates the algorithm using a rotating Kerr black hole. 
Section~\ref{sec:EdGB} applies the algorithm to sGB gravity and derives the results described above. 
Section~\ref{sec:props} constructs a fitted analytical model from the numerical solutions and compares physical observables determined by the numerical solutions and the fits. 
Finally, Section~\ref{sec:concs} summarizes our results and points to future directions. 
For the remainder of this paper we use the following conventions: Greek letters denote spacetime indices; the metric has the spacetime signature $\lb -,+,+,+ \rb$; we use geometric units where $G = 1 = c$.

\section{Numerical Methods}
\label{sec:NM}

The numerical infrastructure extends recent work from~\cite{PhysRevD.101.044024} to axial symmetry, following the approach in~\cite{PhysRevD.93.044047,SCHONAUER1989279}, to build a partial differential equation solver for rotating black hole solutions in an arbitrary modified theory of gravity. The infrastructure uses a relaxed Newton-Raphson method to solve the discretized partial differential equations in two dimensions. The field equations are discretized using a Newton interpolation polynomial which naturally introduces discretization errors that must be controlled. As most of the foundations of this infrastructure is detailed in~\cite{PhysRevD.101.044024}, we will focus on the extensions to axisymmetry in this section.

In axisymmetry, the Newton interpolation polynomial and the discretization error remains identical to the spherical symmetry case but with an additional dimension. Specifically, we replace each $\frac{\pd u}{\pd x}$ and $\frac{\pd u}{\pd y}$ operator with their discretized equivalents which introduces their respective discretization errors
\bal
\bsp
\label{eq:pedx}
\pd_{x} \vec{\bm{e}}_{(x,d)} ={}& \pd_{x} \vec{\bm{u}}_d^{(r+2)} - \pd_{x} \vec{\bm{u}}_d^{(r)}, \\
\pd_{xx} \vec{\bm{e}}_{(x,d)} ={}& \pd_{xx} \vec{\bm{u}}_d^{(r+2)} - \pd_{xx} \vec{\bm{u}}_d^{(r)}, \\
\esp
\end{align}
and
\bal
\bsp
\label{eq:pedy}
\pd_{y} \vec{\bm{e}}_{(y,d)} ={}& \pd_{y} \vec{\bm{u}}_d^{(r+2)} - \pd_{y} \vec{\bm{u}}_d^{(r)}, \\
\pd_{yy} \vec{\bm{e}}_{(y,d)} ={}& \pd_{yy} \vec{\bm{u}}_d^{(r+2)} - \pd_{yy} \vec{\bm{u}}_d^{(r)}. \\
\esp
\end{align}
where $\vec{\bm{e}}_{(x,d)}$ and $\vec{\bm{e}}_{(y,d)}$ are the discretized error vectors\footnote{In this paper, the word \emph{vector} stands for a standard Euclidean vector in flat space.} at each grid point in the $x$ and $y$ dimension respectively. $\vec{\bm{u}}_d^{(r)}$ is the discretized solution vector $\vec{\bm{u}}$ that we wish to minimize to the Newton polynomial order $r$. With two discretization errors now, we obtain an additional discretization error correction equation that must be minimized
\be
{\bm{\mcl{J}}} \Delta \vec{\bm{u}}_{y,e} = - \vec{\bm{D}}_{y,e},
\ee
where ${\bm{\mcl{J}}}$ is the Jacobian matrix, $\Delta \vec{\bm{u}}_{y,e}$ is the correction on the solution vector $\vec{\bm{u}}$ due to the discretization error vector $\vec{\bm{D}}_{y,e}$ in the $y$-dimension. This additional equation must be minimized along with our two previous equations
\bal
{\bm{\mcl{J}}} \Delta \vec{\bm{u}}_{d} &= - \vec{\bm{b}}_{d}, \\
{\bm{\mcl{J}}} \Delta \vec{\bm{u}}_{x,e} &= - \vec{\bm{D}}_{x,e}.
\end{align}
To control the discretization error, we require that the relative correction due to both the $x$ and $y$-dimension ($\vec{\bm{u}}_{x,e}$ and $\vec{\bm{u}}_{y,e}$) discretization error is below a specified tolerance,
\bal
\frac{\left\Vert \Delta \vec{\bm{u}}_{x,e} \right\Vert}{\left\Vert \vec{\bm{u}}_{d} \right\Vert} \leq \mrm{tol}, \\
\frac{\left\Vert \Delta \vec{\bm{u}}_{y,e} \right\Vert}{\left\Vert \vec{\bm{u}}_{d} \right\Vert} \leq \mrm{tol}.
\end{align}

The main addition from spherical to axisymmetry is the treatment of mixed derivatives. We utilize the method in~\cite{SCHONAUER1989279} and treat each mixed derivative as a separate field equation. Namely, we define a new auxiliary variable
\be
\label{eq:mixed-def}
\frac{\pd u_0}{\pd y} = u_1,
\ee
and substitute it into each mixed derivative operator,
\be
\frac{\pd^2 u_0}{\pd x \pd y} = \frac{\pd u_1}{\pd x}.
\ee
We then treat the auxiliary variable definition of Eq.~\eqref{eq:mixed-def} as a separate field equation whose residual we simultaneously must minimize
\be
\label{eq:mixed-FE}
\frac{\pd u_0}{\pd y} - u_1 = b_1,
\ee
which will double the amount of differential equations we must solve.

The relaxed Newton-Raphson method leads to a system of linear equations that must then be solved using various methods. In the spherically symmetric case, iterative solver methods were comparable but had faster convergence over direct methods. In the axisymmetric case, the field equations become less diagonally dominant and iterative methods fail to successfully accelerate the computation time. Due to this, we find that direct methods once again become the faster method because the size of our linear system is not large enough (on the order of millions of elements) for the iterative methods to accelerate convergence.

\section{Validation}
\label{sec:GR}

We now apply our numerical infrastructure using the method described in the previous section to a stationary rotating black hole in general relativity described by the Kerr metric. Although the solution is known analytically, we can use it to benchmark our numerical infrastructure.

The familiar Einstein-Hilbert action in General Relativity in  vacuum is given by
\be
\label{eq:GRaction}
S_{\GR} = \frac{1}{16 \pi} \int d^4 x \sqrt{-g} \; R \,,
\ee
where $R$ is the Ricci scalar and $g$ is the determinant of the metric $\tn{g}{_\mu_\nu}$. Varying the action with respect to the metric gives the vacuum Einstein field equations
\be
\label{eq:GREE}
\tn{G}{_\mu_\nu} = 0 \,,
\ee
where $G_{\mu \nu}$ is the Einstein tensor. 

We begin with an axisymmetric and stationary metric ansatz in isotropic coordinates,\footnote{Note that this is a slightly modified ansatz from~\cite{PhysRevD.101.044024}. This ansatz produces field equations that are easier to diagonalize as we will see later.}
\bal
\begin{split}
ds^2 &= - f (\rho, \theta) d t^2 + \frac{m (\rho, \theta)}{f (\rho, \theta)} \lb d \rho^2 + \rho^2 d \theta^2 \rb \\
+& \frac{l (\rho, \theta)}{f (\rho, \theta)} \rho^2 \ssqth \lb d \phi - \frac{\omega (\rho, \theta)}{\rho} d t \rb^2,
\end{split}
\end{align}
where $\rho$ is the isotropic radial coordinate. For a Kerr metric with mass  $M_0$ and spin $a_0$, the isotropic coordinate $\rho$ is related to the Boyer-Lindquist radial coordinate by
\bal
\begin{split}
r &= \rho \lb 1 + \frac{M_0 + a_0}{2 \rho} \rb \lb 1 + \frac{M_0 - a_0}{2 \rho} \rb, \\
&= \rho + M_0 + \frac{M_0^2 - a_0^2}{4 \rho}.
\end{split}
\end{align}
%
It is convenient to replace the spin parameter $a_0$ with the event horizon radius $\rho_{\Hz}$ using the relation
\be
\label{eq:atorH}
a_0 \equiv \sqrt{M_0^2 - 4 \rho_{\Hz}^2}.
\ee
Replacing this in the above coordinate transformation yields
\be
r = \rho + M_0 + \frac{\rho_{\Hz}^2}{\rho}.
\ee
The Kerr metric in isotropic coordinates is
\bal
\begin{split}
f_{\GR} &= \lb 1 - \frac{\rho_{\Hz}^2}{\rho^2} \rb^2 \frac{F_1}{F_2}, \\
m_{\GR} &= \lb 1 - \frac{\rho_{\Hz}^2}{\rho^2} \rb^2 \frac{F_1^2}{F_2}, \\
l_{\GR} &= \lb 1 - \frac{\rho_{\Hz}^2}{\rho^2} \rb^2, \\
\omega_{\GR} &= \frac{F_3}{F_2},
\end{split}
\end{align}
where
\bal
\begin{split}
\label{eq:F123def}
F_1 &= \frac{2 M_0^2}{\rho^2} + \lb 1 - \frac{\rho_{\Hz}^2}{\rho^2} \rb^2 + \frac{2 M_0}{\rho} \lb 1 + \frac{\rho_{\Hz}^2}{\rho^2} \rb \\
&- \frac{M_0^2 - 4 \rho_{\Hz}^2}{\rho^2} \ssqth, \\
F_2 &= \lsb \frac{2 M_0^2}{\rho^2} + \lb 1 - \frac{\rho_{\Hz}^2}{\rho^2} \rb^2 + \frac{2 M_0}{\rho} \lb 1 + \frac{\rho_{\Hz}^2}{\rho^2} \rb \rsb^2 \\
&- \lb 1 - \frac{\rho_{\Hz}^2}{\rho^2} \rb^2 \frac{M_0^2 - 4 \rho_{\Hz}^2}{\rho^2} \ssqth, \\
F_3 &= \frac{2 M_0 \sqrt{M_0^2 - 4 \rho_{\Hz}^2} \lb 1 + \frac{M_0}{\rho} + \frac{\rho_{\Hz}^2}{\rho^2} \rb}{\rho^2}.
\end{split}
\end{align}
\begin{figure*}[htb]
\begin{center}
\resizebox{8cm}{!}{\includegraphics{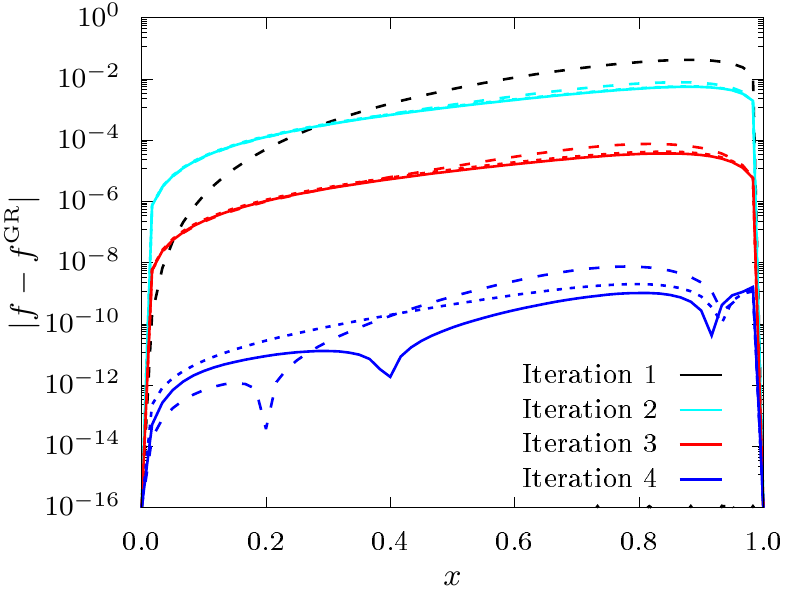}}
\resizebox{8cm}{!}{\includegraphics{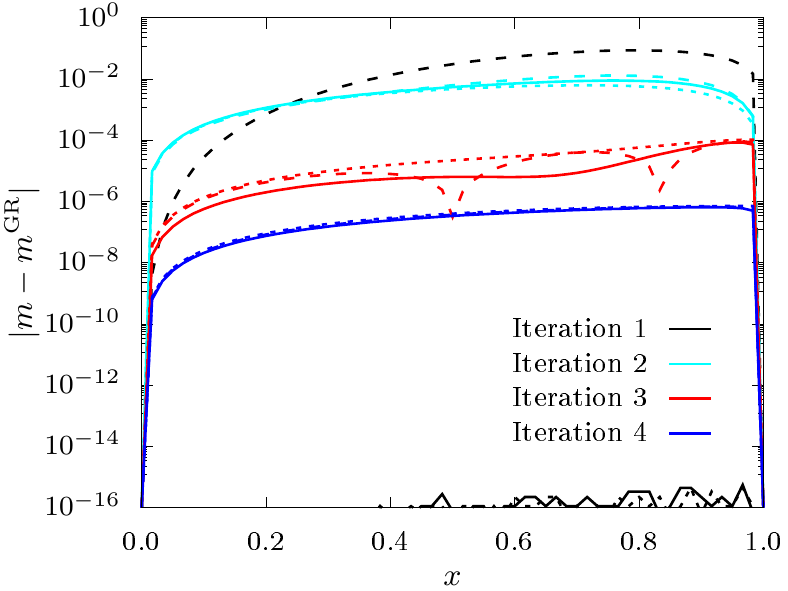}}
\\
\resizebox{8cm}{!}{\includegraphics{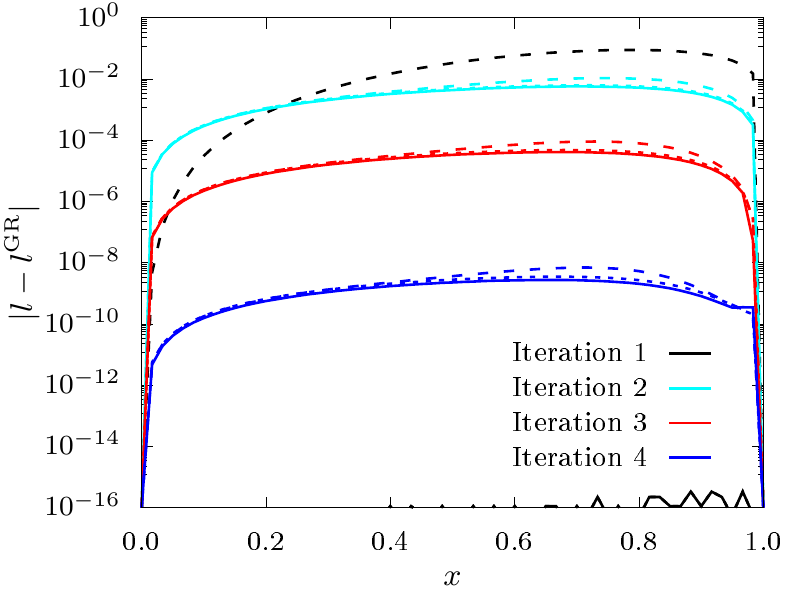}}
\resizebox{8cm}{!}{\includegraphics{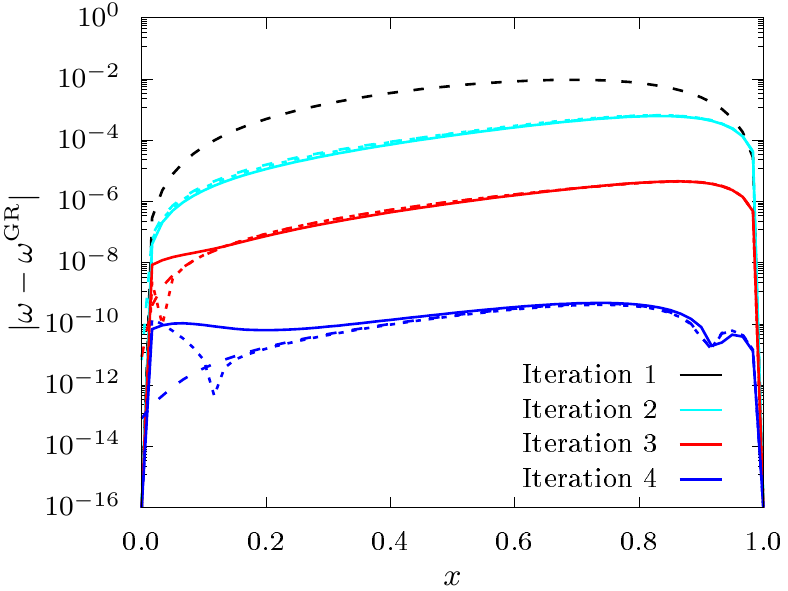}}
\caption{\label{fig:KERRerror} (Color Online) 
(Color online)
Absolute error during each iteration (colored) for each metric element to the Kerr solution for three selected angles. Here we show the metric components $f$ (top left), $m$ (top right), $l$ (bottom left) and $\omega$ (bottom right) for each iteration denoted by color and for three angles $\theta = 0,\pi/4, \pi/2$ denoted by the dotted, dashed, and solid lines respectively. We find that with our chosen initial guess, our numerical infrastructure converges to the Kerr solution to a maximum absolute error of $\mcl{O}(10^{-6})$ and a minimum error of $\mcl{O}(10^{-10})$ in 4 iterations.}
\end{center}
\end{figure*}

The following properties of the Kerr metric in isotropic coordinates are used. On the event horizon, $f_{\GR} |_{\rho = \rho_{\Hz}} = m_{\GR} |_{\rho = \rho_{\Hz}} = l_{\GR} |_{\rho = \rho_{\Hz}} = 0$,  the frame dragging term $\omega$ is a constant
\be
\omega |_{\rho = \rho_{\Hz}} = \omega_{\Hz} = \frac{\rho_{\Hz} \sqrt{M_0^2 - 4 \rho_{\Hz}^2}}{2 M_0 \lb M_0 + 2 \rho_{\Hz} \rb},
\ee
and is proportional to the angular velocity of the black hole event horizon $\Omega_{\Hz}$,
\be
\label{eq:OmegaH}
\Omega_{\Hz} = \frac{\omega_{\Hz}}{\rho_{\Hz}} = \frac{\sqrt{M_0^2 - 4 \rho_{\Hz}^2}}{2 M_0 \lb M_0 + 2 \rho_{\Hz} \rb}.
\ee
In the non-rotating (Schwarzschild) limit $M_0 = 2 \rho_{\Hz}$ and
\bal
\begin{split}
f_{\GR}^{\SH} &= \frac{\lb 1 - \frac{\rho_{\Hz}}{\rho}\rb^2}{\lb 1 + \frac{\rho_{\Hz}}{\rho}\rb^2}, \\
m_{\GR}^{\SH} &= \lb 1 - \frac{\rho_{\Hz}}{\rho}\rb^2 \lb 1 + \frac{\rho_{\Hz}}{\rho}\rb^2, \\
l_{\GR}^{\SH} &= m_{\GR}^{\SH}, \\
\omega_{\GR}^{\SH} &= 0.
\end{split}
\end{align}

Regularity of the solutions along the symmetry axis $\theta = 0$ and $\theta = \pi$ implies that the metric functions should satisfy the boundary conditions 
\bal
\begin{split}
\frac{\pd f}{\pd \theta} |_{\theta = 0,\pi} &= 0, \\
\frac{\pd m}{\pd \theta} |_{\theta = 0,\pi} &= 0, \\
\frac{\pd l}{\pd \theta} |_{\theta = 0,\pi} &= 0, \\
\frac{\pd \omega}{\pd \theta} |_{\theta = 0,\pi} &= 0,
\end{split}
\end{align}
which our solution indeed satisfies. As expected, the metric is asymptotically flat, $f_{\GR} |_{\rho \rarr \infty} = m_{\GR} |_{\rho \rarr \infty} = l_{\GR} |_{\rho \rarr \infty} = 1$ and $\omega_{\GR} |_{\rho \rarr \infty} = 0$. Asymptotically far from the black hole, the observable mass and angular momentum can be extracted from the decay of the metric components
\bal
\begin{split}
\tn{g}{_t_t} &= -f + \frac{l}{f} \omega^2 \ssqth = -1 + \frac{2 M}{\rho} + \mcl{O}\lb \frac{1}{\rho^2} \rb, \\
\tn{g}{_t_\phi} &= -\frac{l}{f} \omega \rho \ssqth = - \frac{2 J}{\rho} \ssqth + \mcl{O}\lb \frac{1}{\rho^2} \rb,
\end{split}
\end{align}
where $M$ and $J$ are the Arnowit-Deser-Misner (ADM) mass and angular momentum respectively. For the Kerr solution, we find $M_{\GR} = M_0$ and $J_{\GR} = M_{0} a_{0} = M_0 \sqrt{M_0^2 - 4 \rho_{\Hz}^2}$.

With this ansatz, we can compute the components of the Einstein tensor $\tn{G}{_\mu_\nu}$. To simplify the partial differential equations, following~\cite{PhysRevD.93.044047}, we use linear combinations of the Einstein tensor to diagonalize the equations with respect to the operator $\mcl{\hat{O}} = \frac{\pd^2}{\pd \rho^2} + \frac{1}{\rho^2} \frac{\pd^2}{\pd \theta^2}$,
\bal
\begin{split}
\label{eq:diagEE}
\frac{m}{f} \lb \tn{G}{^\mu_\mu} - 2 \tn{G}{^t_t} - \frac{2 \omega}{\rho} \tn{G}{^t_\phi} \rb &= \frac{1}{f} \mcl{\hat{O}} f + \ldots, \\
2 \frac{m}{f} \lb \tn{G}{^\phi_\phi} - \frac{\omega}{\rho} \tn{G}{^t_\phi} \rb &= \frac{1}{m} \mcl{\hat{O}} m + \ldots, \\
2 \frac{m}{f} \lb \tn{G}{^\rho_\rho} + \tn{G}{^\theta_\theta} \rb &= \frac{1}{l} \mcl{\hat{O}} l + \ldots, \\
2 \frac{f \, m}{l \, \ssqth} \lb - \frac{1}{\rho} \tn{G}{^t_\phi} \rb &= \mcl{\hat{O}} \omega + \ldots.
\end{split}
\end{align}

As in the spherical symmetry case, we again use a compactified coordinate defined by
\be
\label{eq:xdef}
x = 1 - \frac{\rho_{\Hz}}{\rho}.
\ee
This changes our domain of integration from $\rho \in [\rho_{\Hz}, \infty)$ to the finite domain $x \in [0,1]$. In these compactified isotropic coordinates, the functions have the form 
%
\bal
\begin{split}
\label{eq:KERRsol}
f_{\GR} &= x^2 \lb x-2 \rb^2 \frac{F_1^{\xT}}{F_2^{\xT}}, \\
m_{\GR} &= x^2 \lb x-2 \rb^2 \frac{\lb F_1^{\xT} \rb^2}{F_2^{\xT}}, \\ 
l_{\GR} &= x^2 \lb x-2 \rb^2, \\ 
\omega_{\GR} &= \frac{F_3^{\xT}}{F_2^{\xT}},
\end{split}
\end{align}
where $F_1^{\xT}, F_2^{\xT}$, and $F_3^{\xT}$ are the functions from Eq.~\eqref{eq:F123def} in compactified coordinates. As before, we have similar boundary conditions, $f_{\GR} |_{x=0} = m_{\GR} |_{x=0} = l_{\GR} |_{x=0} = 0$ and $\omega_{\GR} |_{x=0} = \omega_{\Hz}$. At infinity we have $f_{\GR} |_{x=1} = m_{\GR} |_{x=1} = l_{\GR} |_{x=1} = 1$ and $\omega_{\GR} |_{x=1} = 0$.

To prepare our field equations for numerical integration, we make an additional substitution following~\cite{PhysRevD.57.6138}. We find that this substitution is necessary to eliminate a numerical divergence on the event horizon in the scalar Gauss-Bonnet case considered in Sec.~\ref{sec:EdGB}. We replace the metric functions with corresponding barred functions defined by
\bal
\begin{split}
\label{eq:bardef}
f &= x^2 \bar{f}, \\
m &= x^2 \bar{m}, \\
l &= x^2 \bar{l},
\end{split}
\end{align}
which removes this numerical divergence. This substitution leaves the boundary conditions as $x \rarr 1$ unchanged. At the horizon, the boundary conditions are obtained from examining an expansion of the metric functions around $x=0$ (see~\cite{PhysRevD.57.6138}) and become
\bal
\begin{split}
\lb \bar{f} - \frac{\pd \bar{f}}{\pd x} \rb |_{x=0} &= 0, \\
\lb \bar{m} + \frac{\pd \bar{m}}{\pd x} \rb |_{x=0} &= 0, \\
\lb \bar{l} + \frac{\pd \bar{l}}{\pd x} \rb |_{x=0} &= 0.
\end{split}
\end{align}

Similar to the spherically symmetric case, the Newton-Raphson method requires an initial guess for the numerical system. We shall again, choose an initial guess that is a small perturbation away from the Kerr metric and that satisfies the boundary conditions
\bal
\begin{split}
\label{eq:uguess}
u_0^{(0)} &= \bar{f}_{\GR} \lsb 1 + \delta \, \Delta x \Delta y \rsb, \\
u_1^{(0)} &= \bar{m}_{\GR} \lsb 1 + \delta \, \Delta x \Delta y \rsb, \\
u_2^{(0)} &= \bar{l}_{\GR} \lsb 1 + \delta \, \Delta x \Delta y \rsb, \\
u_3^{(0)} &= \omega_{\GR} \lsb 1 + \delta \, \Delta x \Delta y \rsb,
\end{split}
\end{align}
where $\delta = 0.1$ and can be adjusted to improve or worsen the initial guess.\footnote{We find that the convergence in GR is largely independent of the value of $\delta$. Even initial guess values as large as $\delta=1$ converge to the desired solution in less than 10 iterations.} The normalized functions $\Delta x$ and $\Delta \theta$ are chosen to be
\bal
\begin{split}
\Delta x &= \frac{256}{27} x^3 \lb 1 - x \rb, \\
\Delta y &= \frac{512}{\pi^3} \lb \frac{\theta}{\pi/2} \rb^3 \lb 1 - \frac{\theta}{\pi/2} \rb^3.
\end{split}
\end{align}

To solve our problem numerically, we begin by replacing the metric functions of our ansatz with their barred definitions of Eq.~\eqref{eq:bardef}. We then define the auxiliary mixed derivative functions
\bal
\begin{split}
\label{eq:umixed-def}
\frac{\pd \bar{f}}{\pd \theta} &\equiv u_4, \\
\frac{\pd \bar{m}}{\pd \theta} &\equiv u_5, \\
\frac{\pd \bar{l}}{\pd \theta} &\equiv u_6, \\
\frac{\pd \omega}{\pd \theta} &\equiv u_7,
\end{split}
\end{align}
and replace each mixed derivative operator given by\footnote{We find that it is unnecessary to make the second order replacement $\frac{\pd^2 \bar{f}}{\pd \theta^2} = \frac{\pd u_4}{\pd \theta}$ as the second derivative $\frac{\pd^2 \bar{f}}{\pd \theta^2}$ terms can be evaluated very accurately with our Newton polynomial representation. We find that this substitution only slows down convergence.}
\bal
\begin{split}
\label{eq:umixed-sub}
\frac{\pd^2 \bar{f}}{\pd x \pd \theta} &= \frac{\pd u_4}{\pd x}, \\
\frac{\pd^2 \bar{m}}{\pd x \pd \theta} &= \frac{\pd u_5}{\pd x}, \\
\frac{\pd^2 \bar{l}}{\pd x \pd \theta} &= \frac{\pd u_6}{\pd x}, \\
\frac{\pd^2 \omega}{\pd x \pd \theta} &= \frac{\pd u_7}{\pd x},
\end{split}
\end{align}
in the diagonalized Einstein equations of Eq.~\eqref{eq:diagEE} in compactified isotropic coordinates. From Eq.~\eqref{eq:mixed-FE}, the mixed derivative definitions above add 4 additional field equations we must solve simultaneously with the Einstein equations and we obtain a nonlinear system of 8 partial differential equations for our 8 functions to solve: $\bar{f}, \bar{m}, \bar{l}, \omega, u_4, u_5, u_6, u_7$.

We then discretize our differential operators using their Newton polynomial representation of order $r=16$ on a 2-dimensional grid of $61 \times 31$ points and initialize our solver with the initial guess of Eq.~\eqref{eq:uguess}. The two input parameters that we must specify is the horizon radius where we choose $\rho_{\Hz}$ and the angular velocity on the event horizon $\Omega_{\Hz}$. For all computations in this paper, we set $\rho_{\Hz} = 1$. The horizon angular velocity is chosen to coincide with that of a Kerr black hole of dimensionless spin $\chi^{\GR} = J_{\GR}/M_{0}^{2} = 0.6$ which from Eq.~\eqref{eq:OmegaH} implies $\Omega_{\Hz} = 0.0667$, where we have set $\rho_{\Hz} = 1$.
We find that our numerical infrastructure converges to the desired solution below our specified tolerance of $\mrm{tol} = 10^{-5}$ in 4 iterations. The absolute error between the metric functions and the Kerr solution for each iteration is shown in Fig.~\ref{fig:KERRerror}. This figure validates our numerical code to construct stationary and axisymmetric black hole solutions.

\section{Axially Symmetric Black Holes in Scalar-Gauss-Bonnet Gravity}
\label{sec:EdGB}

In this section we solve the modified Einstein field equations in sGB gravity with both a linear coupling and an exponential coupling function, assuming a vacuum spacetime that is stationary and axially symmetric.

\subsection{Action and Field equations}

The action in scalar-Gauss-Bonnet gravity in vacuum is given by
\be
\label{eq:action}
S = \frac{1}{16 \pi} \int d^4 x \sqrt{-g} \lsb R - \tn{\nabla}{_\mu} \psi \tn{\nabla}{^\mu} \psi + 2 \, \alpha F(\psi) \mcl{G} \rsb,
\ee
where $R$ is the Ricci scalar and $g$ is the determinant of the metric $\tn{g}{_\mu_\nu}$. The real dimensionless scalar field $\psi$ is coupled to the Gauss-Bonnet invariant
\be
\mcl{G} = R^2 - 4 \tn{R}{^\mu^\nu} \tn{R}{_\mu_\nu} + \tn{R}{^\mu^\nu^\rho^\sigma} \tn{R}{_\mu_\nu_\rho_\sigma},
\ee
through a function of the scalar field $F(\psi)$ with a coupling constant $\alpha$ that has dimensions of length squared.

By varying the action with respect to the metric and the scalar field we obtain two field equations. Variation with respect to the metric field yields
\be
\label{eq:EE}
\tn{G}{_\mu_\nu} - \tn{T}{_\mu_\nu} + \alpha \, \tn{K}{_\mu_\nu} = 0,
\ee
where the scalar field stress-energy tensor is
\be
\label{eq:T}
\tn{T}{_\mu_\nu} = \tn{\cd}{_\mu} \psi \tn{\cd}{_\nu} \psi - \frac{1}{2} \tn{g}{_\mu_\nu} \tn{\cd}{^\gamma} \psi \tn{\cd}{_\gamma} \psi,
\ee
and
\bal
\begin{split}
\label{eq:Kdef}
\tn{K}{_\mu_\nu} &= \lb \tn{g}{_\rho_\mu} \tn{g}{_\delta_\nu} + \tn{g}{_\rho_\nu} \tn{g}{_\delta_\mu} \rb \times \\
& \tn{\cd}{_\sigma} \lsb \tn{\epsilon}{^\gamma^\delta^\alpha^\beta} \tn{\epsilon}{^\rho^\sigma^\lambda^\eta} \tn{R}{_\lambda_\eta_\alpha_\beta} \tn{\cd}{_\gamma} F(\psi) \rsb.
\end{split}
\end{align}
Variation with respect to the scalar field yields
\be
\label{eq:box}
\Box \psi + \alpha \, \frac{\pd F}{\pd \psi} \mcl{G} = 0.
\ee
The scalar field is subject to the following boundary conditions:  it must be asymptotically flat, and its first derivative must vanish on the horizon in isotropic coordinates, which follows from the regularity condition on the horizon~\cite{PhysRevD.54.5049, Sotiriou:2013qea,PhysRevD.90.124063}, namely
\be
\label{eq:psiBC}
\frac{\pd \psi}{\pd \rho} |_{\rho \rarr \rho_{\Hz}} = 0, \qquad \psi |_{\rho \rarr \infty} = 0.
\ee

In this paper we will consider two coupling functions typically explored in sGB gravity,
\begin{align}
\begin{split}
F(\psi) &= \psi \; \; \, \leftrightarrow \;  {\rm{linear}} \; {\rm{sGB}}\,,
\\
F(\psi) &= e^{\psi} \; \leftrightarrow \; {\rm{EdGB}}\,,
\end{split}
\end{align} 
and will consider them separately in the following sections.  

\subsection{Linear Scalar-Gauss-Bonnet Gravity}

\begin{figure}[h]
\begin{center}
\resizebox{8cm}{!}{\includegraphics{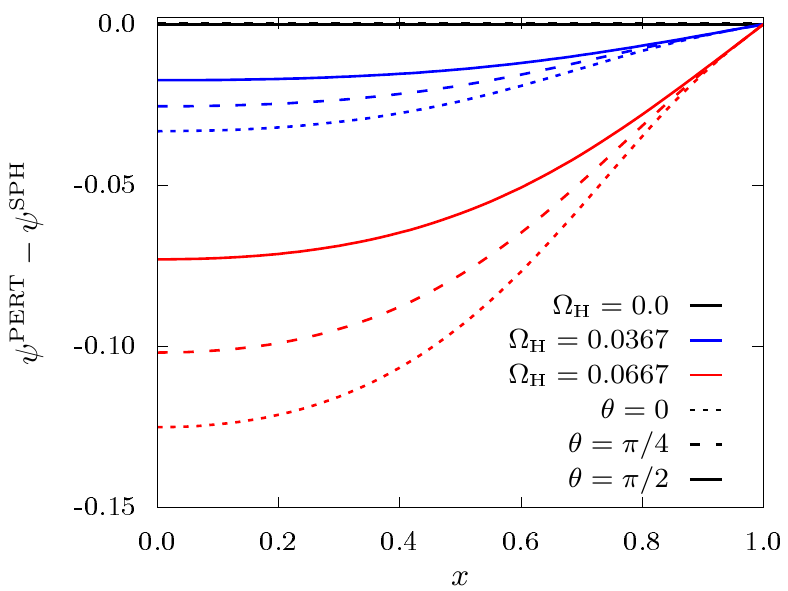}}
\caption{\label{fig:PERT_SPHpsi} 
(Color online)
Difference between the axisymmetric small coupling expansion and the analytic perturbative spherically symmetric solution for the scalar field for three selected angles. The three event horizon angular velocity values, $\Omega_{\Hz} = 0.0,0.0367$, and $0.0667$ are denoted by the black, blue, and red colors respectively and the three angles $\theta = 0,\pi/4, \pi/2$ are denoted by the dotted, dashed, and solid lines respectively.}
\end{center}
\end{figure}
\begin{figure*}[htb]
\begin{center}
\includegraphics[width=7.5cm,clip=true]{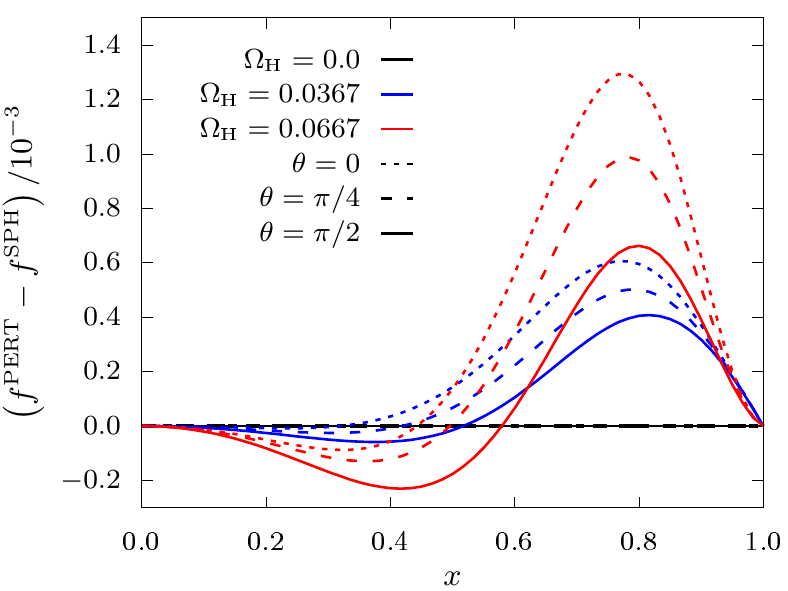}
\includegraphics[width=7.5cm,clip=true]{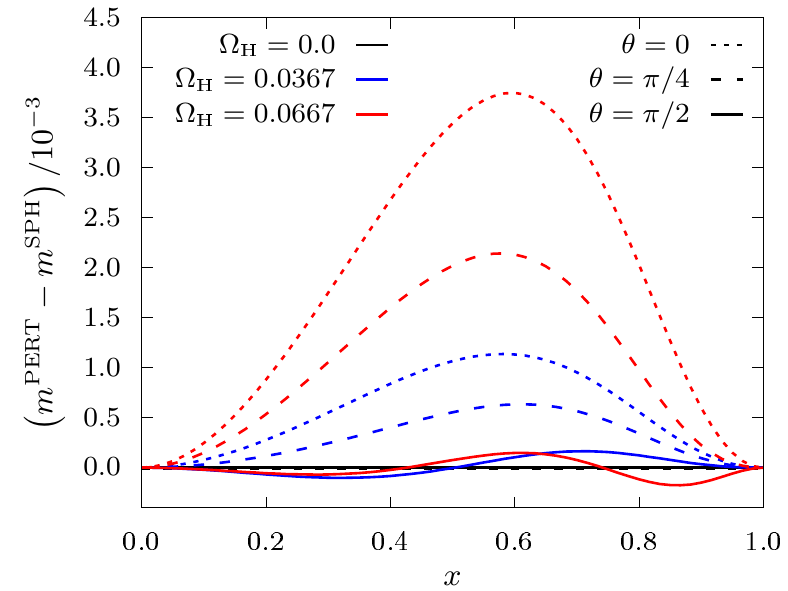}
\\
\includegraphics[width=7.5cm,clip=true]{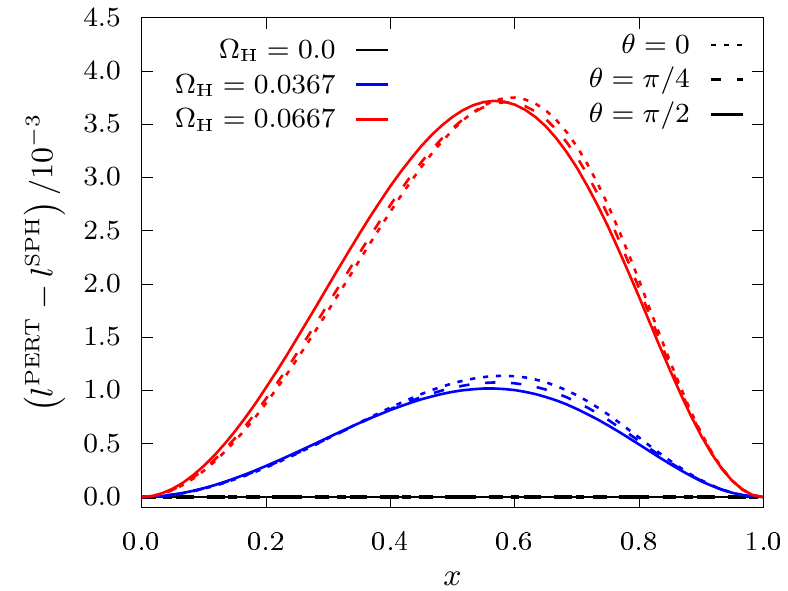}
\includegraphics[width=7.5cm,clip=true]{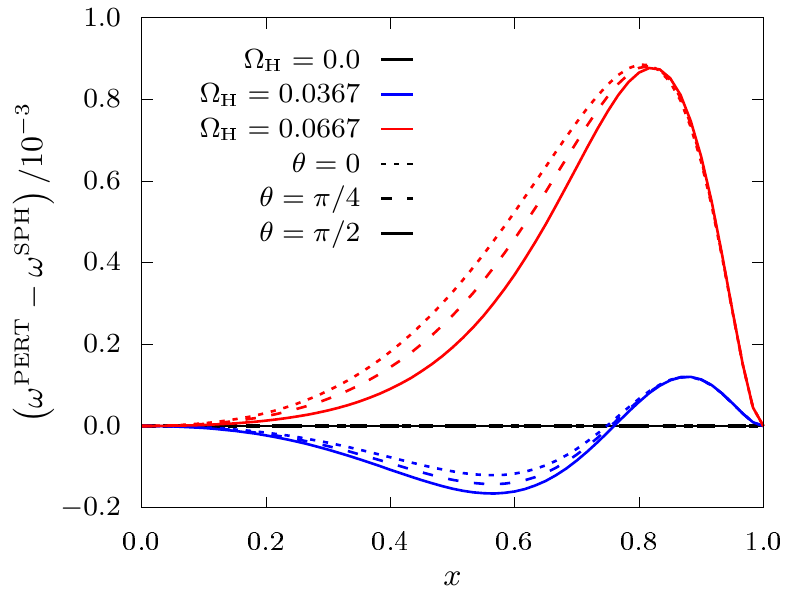}
\caption{\label{fig:PERT_SPHg} 
(Color online)
Difference between metric elements for axisymmetric small coupling expansion and the analytic perturbative spherically symmetric solution for three selected angles. Here we show the metric components $f$ (top left), $m$ (top right), $l$ (bottom left) and $\omega$ (bottom right) for three event horizon angular velocity values, $\Omega_{\Hz} = 0.0,0.0367$, and $0.0667$ denoted by the black, blue, and red colors respectively and for three angles $\theta = 0,\pi/4, \pi/2$ denoted by the dotted, dashed, and solid lines respectively. }
\end{center}
\end{figure*}

Let us first consider solving the field equations for an axially symmetric black hole \emph{perturbatively} in the coupling $\alpha$. If we assume the dimensionless coupling $\bar{\alpha} \equiv \alpha/\rho_{\Hz}^2 \ll 1$ where $\rho_{\Hz}$ sets the order of the curvature length of the system, we can perturbatively expand our metric as
\be
\tn{g}{_\mu_\nu} = \tnst{g}{^{(0)}_\mu_\nu} + \epsilon \tnst{g}{^{(1)}_\mu_\nu} + \epsilon^2 \tnst{g}{^{(2)}_\mu_\nu},
\ee
where $\epsilon \ll 1$ is a bookkeeping parameter and $\alpha = \mcl{O}(\epsilon)$. This expansion with our metric ansatz is,
\bal
\begin{split}
f &= f_0 + \epsilon f_1 + \epsilon^2 f_2, \\
m &= m_0 + \epsilon m_1 + \epsilon^2 m_2, \\
l &= l_0 + \epsilon l_1 + \epsilon^2 l_2, \\
\omega &= \omega_0 + \epsilon \omega_1 + \epsilon^2 \omega_2, \\
\psi &= \psi_0 + \epsilon \psi_1 + \epsilon^2 \psi_2,
\end{split}
\end{align}
 We can then substitute this ansatz into our field equations and expand order by order in $\epsilon$.

To $\mcl{O}(\epsilon^0)$, we find 
\bal
\begin{split}
\label{eq:EEO0}
\tnst{G}{^{(0)}_\mu_\nu} - \tnst{T}{^{(0)}_\mu_\nu} &= 0, \\
\tn{\Box}{^{(0)}} \tn{\psi}{^{(0)}} &= 0,
\end{split}
\end{align}
where $\tnst{G}{^{(0)}_\mu_\nu}, \tnst{T}{^{(0)}_\mu_\nu}$, and $\tn{\Box}{^{(0)}}$ are the Einstein tensor, scalar field stress-energy tensor, and the d'Alambertian associated with the background metric $\tnst{g}{^{(0)}_\mu_\nu}$. By requiring the scalar field be asymptotically flat and regular on the horizon, we find $\tn{\psi}{^{(0)}} = 0$ which implies that $\tnst{T}{^{(0)}_\mu_\nu} = 0$. As expected we then see that $\tnst{g}{^{(0)}_\mu_\nu}$ is the solution to $\tnst{G}{^{(0)}_\mu_\nu} = 0$ which is the Kerr metric and each $f_0,m_0,l_0,\omega_0$ correspond to their respective Kerr values from Eq.~\eqref{eq:KERRsol}.  Indeed, this is expected by the well-known no hair theorem that covers the case of a minimally coupled scalar field \cite{Hawking:1972qk}. 

At $\mcl{O}(\epsilon)$, we find
\bal
\begin{split}
\label{eq:EEO1}
\tnst{G}{^{(1)}_\mu_\nu} - \tnst{T}{^{(1)}_\mu_\nu} + \alpha \,\tnst{K}{^{(0)}_\mu_\nu} &= 0, \\
\tn{\Box}{^{(1)}} \tn{\psi}{^{(0)}} + \tn{\Box}{^{(0)}} \tn{\psi}{^{(1)}} + \alpha \, \tn{\mcl{G}}{^{(0)}} &= 0. 
\end{split}
\end{align}
Since $\tn{\psi}{^{(0)}} = 0$ from before, we know $\tnst{K}{^{(0)}_\mu_\nu} = 0$. Additionally, $\tnst{T}{^{(1)}_\mu_\nu} = 0$ because the stress-energy tensor is $\mcl{O}(\psi^2)$. Thus the metric perturbation at $\mcl{O}(\epsilon)$ vanishes, $\tnst{g}{^{(1)}_\mu_\nu} = 0$ and $f_1=m_1=l_1=\omega_1 = 0$. The scalar field equation then simplifies to
\be
\label{eq:EEpsiO1}
\tn{\Box}{^{(0)}} \tn{\psi}{^{(1)}} + \alpha \, \tn{\mcl{G}}{^{(0)}} = 0. 
\ee
In spherical symmetry, the scalar field correction at this order can be calculated analytically~\cite{PhysRevD.83.104002, Sotiriou:2013qea,PhysRevD.90.124063}, while for axisymmetric backgrounds, it has only been found perturbatively in a slow-rotation expansion~\cite{PhysRevD.90.044066, PhysRevD.84.087501, PhysRevD.92.083014}.

At $\mcl{O}(\epsilon^2)$, the modified field equations are
\bal
\label{eq:EEO2}
\tnst{G}{^{(2)}_\mu_\nu} - \tnst{T}{^{(2)}_\mu_\nu} + \alpha \,\tnst{K}{^{(1)}_\mu_\nu} &= 0, \\
\tn{\Box}{^{(2)}} \tn{\psi}{^{(0)}} + \tn{\Box}{^{(1)}} \tn{\psi}{^{(1)}} + \tn{\Box}{^{(0)}} \tn{\psi}{^{(2)}} + \alpha \, \tn{\mcl{G}}{^{(1)}} &= 0. 
\end{align}
Because $\tnst{g}{^{(1)}_\mu_\nu} = 0$, we know that $\tn{\Box}{^{(1)}} =  \tn{\mcl{G}}{^{(1)}} = 0$ which simplifies the scalar field equation to
\be
\label{eq:EEpsiO2}
\tn{\Box}{^{(0)}} \tn{\psi}{^{(2)}} = 0,
\ee
which implies that $\tn{\psi}{^{(2)}} = 0$ by imposing asymptotic flatness and regularity on the horizon. Thus, the nontrivial modified field equations of interest are Eqs.~\eqref{eq:EEpsiO1} and~\eqref{eq:EEO2}. In spherical symmetry~\cite{PhysRevD.83.104002, Sotiriou:2013qea,PhysRevD.90.124063} and in the slow rotation limit~\cite{CAMPBELL1992199, MIGNEMI1993299, PhysRevD.90.044066, PhysRevD.84.087501, PhysRevD.92.083014}, these equations can be analytically solved order by order because the scalar field equation is sourced by the Gauss-Bonnet invariant evaluated on the unperturbed background. In spherical symmetry, in our compactified coordinate system~\eqref{eq:xdef} the perturbed solution to second order is,
\bw
\bal
\begin{split}
\label{eq:pertsol}
f_2^{\SPH} &= \frac{\alpha^2 x^2 \lb x-1 \rb}{4620 \rho_{\Hz}^4 \lb x-2 \rb^{14}} \bigg[1117 x^{10}-24574 x^{9}+246510 x^8-1415920 x^7+4941728 x^6 \\&-10150448 x^5+11892496 x^4-7411712 x^3+2000768 x^2-98560 x+19712 \bigg], \\
m_2^{\SPH} &= - \frac{8 \alpha^2 x^2 \lb x-1 \rb^2}{1155 \rho_{\Hz}^4 \lb x-2 \rb^{10}} \bigg[71 x^8-1420 x^7+11554 x^6-49788 x^5+118374 x^4-167280 x^3 \\ &+147600 x^2-78720 x+19680 \bigg], \\
l_2^{\SPH} &= m_2^{\SPH}, \\
\omega_2^{\SPH} &= 0, \\
\psi_1^{\SPH} &= \frac{\alpha \lb 1 - x\rb}{3 \rho_{\Hz}^2 \lb x-2 \rb^6} \lsb 3 x^4-30 x^3+118 x^2-176 x+88 \rsb.
\end{split}
\end{align}
\ew

In axial symmetry, using a slow rotation expansion around the dimensionless spin $\chi = \frac{a}{M} \ll 1$, solutions have been found to $\mcl{O}(\alpha^2, \chi^2)$~\cite{PhysRevD.90.044066} and $\mcl{O}(\alpha^{14}, \chi^5)$~\cite{PhysRevD.92.083014}. We cannot directly compare these solutions in the slow rotation limit to the solutions in this work because they are calculated in different coordinate systems. A proper comparison would require calculating the solution to the same order in isotropic coordinates. Instead of doing this, we solve these equations directly without perturbatively expanding in rotation.

To solve Eqs.~\eqref{eq:EEpsiO1} and~\eqref{eq:EEO2} we apply our numerical infrastructure to the partially decoupled nonlinear partial differential equations using the method described in Sec.~\ref{sec:NM}. One could solve for the scalar field first using Eq.~\eqref{eq:EEpsiO1} and then use the result to solve Eq.~\eqref{eq:EEO2} as is done in analytic calculations. However, we find no noticeable difference between solutions obtained this way and solutions obtained by solving both equations simultaneously, which our code can handle. This is possible because the scalar field equation is partially decoupled from the metric perturbation equations, i.e.~the scalar field equation only depends on the known GR background to zeroth order and it converges very rapidly. Each successive iteration then only needs to minimize the metric perturbations. We choose an initial grid of $61 \times 31$ points and a Newton polynomial order $r = 16$. For the actual computation, we set $\rho_{\Hz} = 1$. We set the desired tolerance of the solution to $\mrm{tol} = 10^{-5}$ which is both placed on the residual and on the relative tolerance of the discretization correction. We use the spherically symmetric perturbed corrections of Eq.~\eqref{eq:pertsol} as our initial guess and convergence typically occurs within 1 to 3 iterations. Figure~\ref{fig:PERT_SPHg} compares the numerical perturbed rotating solution to the analytically known spherically symmetric solution. From these plots, we can verify that the perturbative solution in the spherically symmetric limit ($\Omega_{\Hz}=0$) exactly recovers the analytic spherically symmetric solution.

\begin{figure}[h]
\begin{center}
\resizebox{8cm}{!}{\includegraphics{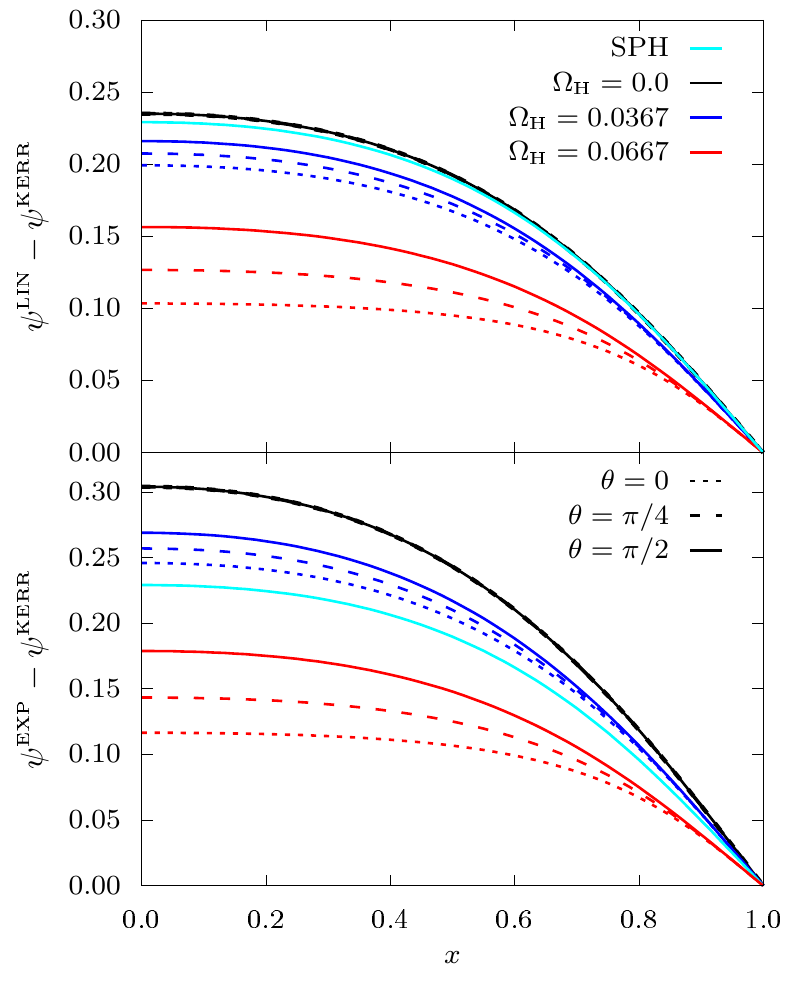}}
\caption{\label{fig:StackKERRpsi} 
(Color online)
The scalar field profile using both the linear coupling (top) and exponential coupling (bottom) for three selected angles. The three event horizon angular velocity values, $\Omega_{\Hz} = 0.0,0.0367$, and $0.0667$ are denoted by the black, blue, and red colors respectively and the three angles $\theta = 0,\pi/4, \pi/2$ are denoted by the dotted, dashed, and solid lines respectively. The analytic perturbative spherically symmetric solution is in cyan. Like for the deviations of the metric functions, we find very good agreement between the nonrotating linear sGB scalar field and the analytic spherically symmetric perturbation but the scalar charge is suppressed for larger spin values.}
\end{center}
\end{figure}

\begin{figure*}[p]
\begin{center}
\resizebox{8cm}{!}{\includegraphics{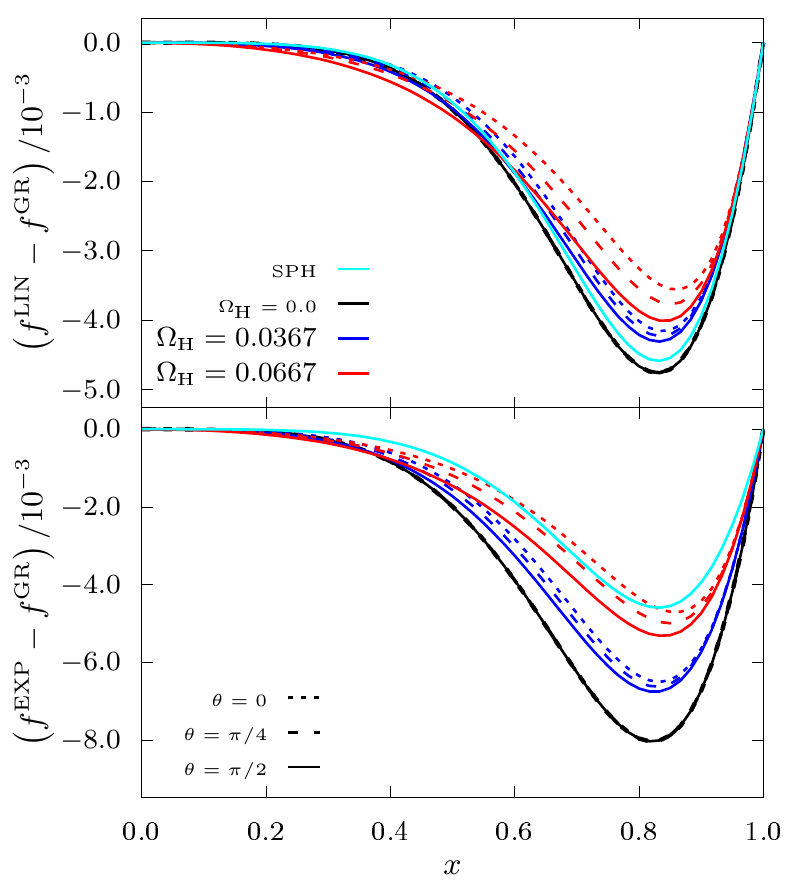}}
\resizebox{8cm}{!}{\includegraphics{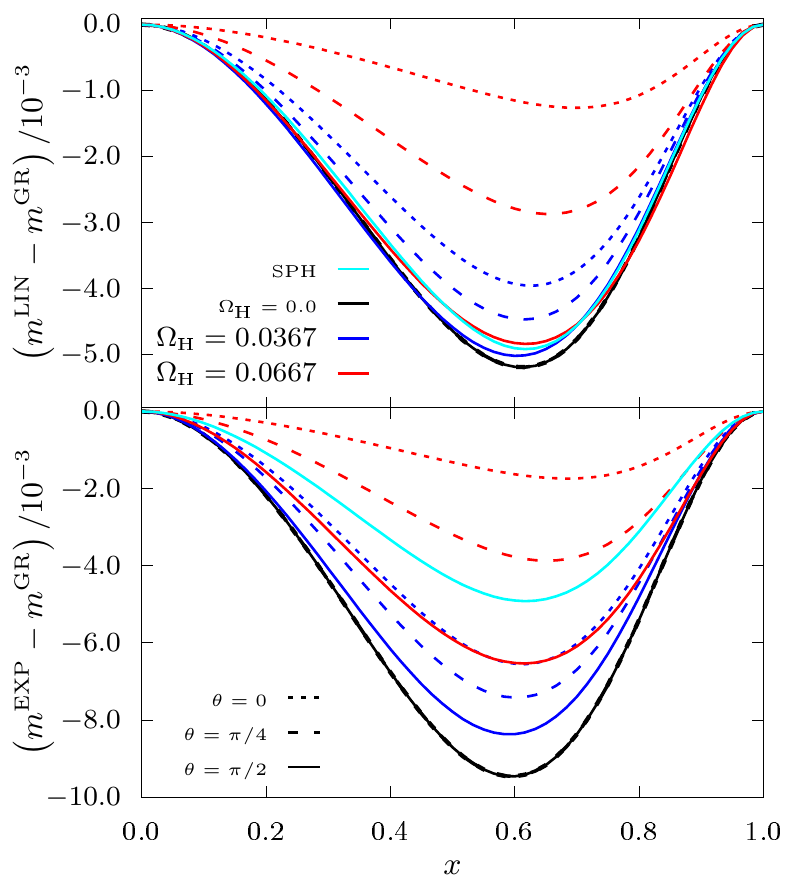}}
\\
\resizebox{8cm}{!}{\includegraphics{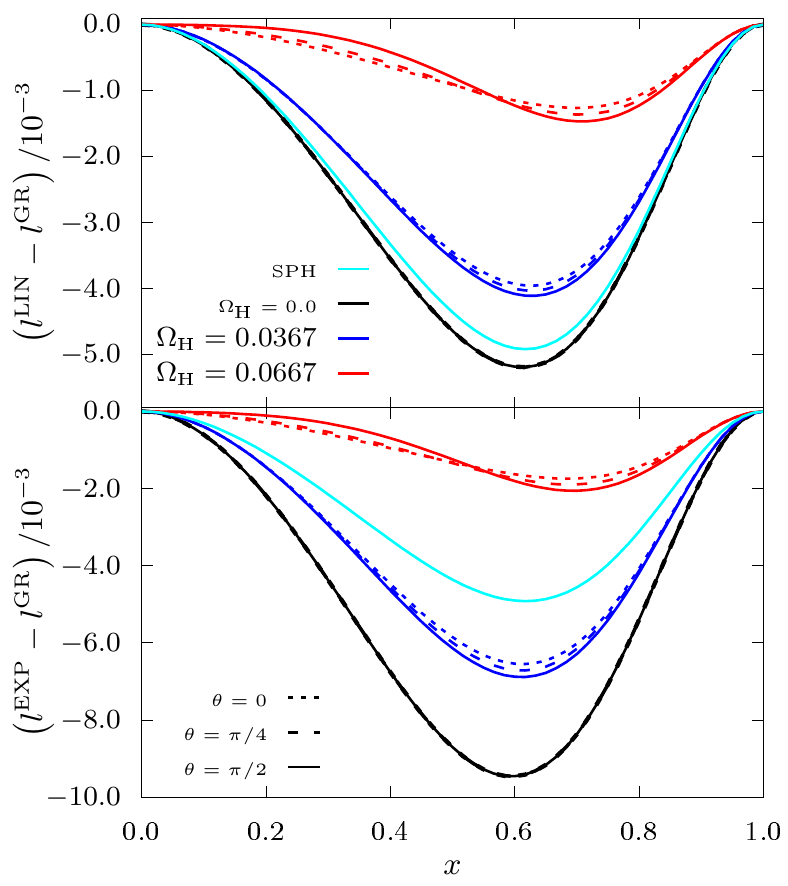}}
\resizebox{8cm}{!}{\includegraphics{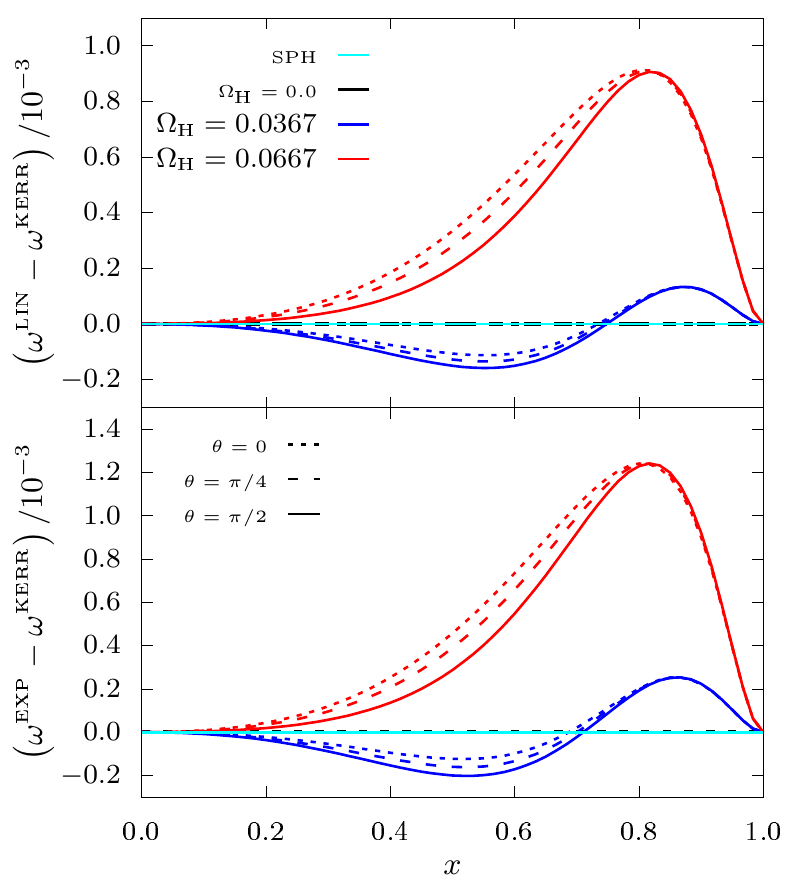}}
\caption{\label{fig:StackKERRg} 
(Color online) Rescaled difference between metric elements and the Kerr metric for both the linear coupling solution (top half of each panel) and the exponential coupling solution (bottom half of each panel) for three selected angles. Here we show the metric components $f$ (top left panel), $m$ (top right panel), $l$ (bottom left panel) and $\omega$ (bottom right panel) for three event horizon angular velocities, $\Omega_{\Hz} = 0.0, 0.0367$, and $0.0667$ denoted by the black, blue, and red colors respectively and for three angles $\theta = 0,\pi/4, \pi/2$ denoted by the dotted, dashed, and solid lines respectively. We also show the analytic perturbative spherically symmetric solutions in cyan. Notice how the nonrotating linear sGB solution has a larger deviation than the analytic perturbative spherically symmetric solution as expected but that this deviation is then suppressed for larger spin values. For the exponential coupling (EDGB) solution we find an even larger deviation from GR than with the linear sGB coupling, as expected.}
\end{center}
\end{figure*}

With this perturbed solution at hand, we can calculate the \emph{full nonlinear} solution to the modified field equations in scalar Gauss-Bonnet gravity. The modified field equations are Eqs.~\eqref{eq:EE} and~\eqref{eq:box} with $F(\psi) = \psi$. In the top of each panel in Figs.~\ref{fig:StackKERRpsi} and~\ref{fig:StackKERRg} we show the difference between the full non-linear sGB solution and the Kerr solution for the scalar field and each metric element respectively for three different angles and three event horizon angular velocities. We recover that the magnitude of the deviation from GR is slightly larger in the full non-linear sGB solution in the spherically symmetric case where $\Omega_{\Hz} = 0$ (solid black line) than in the analytic perturbative spherically symmetric solution (solid cyan line). As we increase the rotation, the deviation from GR decreases as the magnitude of the polar profile (solid-dashed-dotted lines) takes shape. Note that for the $m$ metric function (top right panel) the polar profile on the equator $\theta = \pi/2$ (solid colored lines) remains relatively constant for different values of rotation, while the polar profile at the pole $\theta = 0$ (dotted lines) has enhanced variation in comparison to the other metric elements, whose profiles are relatively similar but change only in magnitude.

Let us point out that the physical dimensionless spin $\chi = a_{0}/M_{0}$ of the black hole will depend on $\bar{\alpha}$. It is for this reason that we report the angular velocity of the event horizon from Eq.~\eqref{eq:OmegaH} to compare our solutions, which is also the input parameter to our numerical infrastructure. Therefore, although each rotating solution represents a rotating black hole with the same event horizon angular velocity, due to their different $\bar{\alpha}$, their physical dimensionless spin $\chi$ will vary slightly. From our range of $\bar{\alpha}$, we find that these differences in $\chi$ are smaller than $2 \%$.

%

\subsection{Einstein-dilaton-Gauss-Bonnet Gravity}

Let us now consider the case of an exponential coupling function. The resulting field equations are Eqs.~\eqref{eq:EE} and~\eqref{eq:box} with $F(\psi) = e^{\psi}$. We find a full non-linear numerical solution using the computational infrastructure of Sec.~\ref{sec:NM}, with the same choices for the grid spacing, Newton polynomial order, etc as in the previous subsection. We show the results in the bottom of each panel in Figs.~\ref{fig:StackKERRpsi} and~\ref{fig:StackKERRg}. This time we find a much larger deviation from GR in the EdGB solutions than in the linear sGB coupling case of the previous subsection by comparing the full non-linear EDGB solution in spherical symmetry where $\Omega_{\Hz} = 0$ (solid black line) to the analytic perturbative spherically symmetric solution (solid cyan line). As we increase the angular momentum of the black hole, the magnitude of the deviation from GR is suppressed. 

We also find a much larger variation of the polar profile for the $m$ metric function than for the $f,l,$ and $\omega$ components. This is particularly interesting, as this metric function happens to have a negligible impact on the physical observables we have calculated. For example, as we will see in the next section, geodesics in an axially symmetric spacetime are completely independent of the $\tn{g}{_r_r}$ component of the metric. Although strictly speaking this is not true for an isotropic metric because $\rho^2 \ssqth \tn{g}{_\phi_\phi} = \tn{g}{_\rho_\rho}$,  these results suggest that even in isotropic coordinates, the dependence on the $\tn{g}{_\rho_\rho}$ metric function is minimal. With our nonlinear numerical solutions at hand, we now use these solutions to construct analytical fitted models and we compare physical observables like the location of the innermost-stable-circular orbit and the light ring.


\section{Properties of Solution}
\label{sec:props}

In this section we explore some physical properties of the numerical solutions found in the previous sections. We begin by finding analytical models that we fit to the data to provide accurate, closed-form expressions that allow for the rapid computation of physical observables. We then use the numerical results to calculate the location of the innermost stable circular orbit (ISCO) and the light ring (LR) by analyzing the motion of null and timelike geodesics. We use a Newton-Raphson method to numerically calculate the location of the ISCO and the LR from the resulting equations. 

\subsection{Fitting Function}

In the compactified coordinate system introduced in Eq.~\eqref{eq:xdef}, the full nonlinear solutions for a given coupling $\bar{\alpha}$ can be expressed as
\bal
\begin{split}
\label{eq:fitstruct}
f (x,\theta) ={}& f_{\GR} + f_{\nonlin}(x,\theta), \\
m (x,\theta) ={}& m_{\GR} + m_{\nonlin}(x,\theta), \\
l (x,\theta) ={}& l_{\GR} + l_{\nonlin}(x,\theta), \\
\omega (x,\theta) ={}& \omega_{\GR} + \omega_{\nonlin}(x,\theta), \\
\psi (x,\theta) ={}& \psi_{\nonlin}(x,\theta). \\
\end{split}
\end{align}
We propose best fit models for the non-linear corrections of the form
\bal
\label{eq:nonfit}
\begin{split}
f_{\nonlin}(x,\theta) ={}& x^2 \lb x-1 \rb \lb \sum_i \sum_j a_{i,j} x^i P_{j}(\cos{\theta}) \rb, \\
m_{\nonlin}(x,\theta) ={}& x^2 \lb x-1 \rb^2 \lb \sum_i \sum_j b_{i,j} x^i P_{j}(\cos{\theta}) \rb, \\
l_{\nonlin}(x,\theta) ={}& x^2 \lb x-1 \rb^2 \lb \sum_i \sum_j c_{i,j} x^i P_{j}(\cos{\theta}) \rb, \\
\omega_{\nonlin}(x,\theta) ={}& \lb x-1 \rb^2 \lb \sum_i \sum_j d_{i,j} x^i P_{j}(\cos{\theta}) \rb, \\
\psi_{\nonlin}(x,\theta) ={}& \lb x-1 \rb \lb \sum_i \sum_j e_{i,j} x^i P_{j}(\cos{\theta}) \rb,
\end{split}
\end{align}
%
where $x^i$ is a polynomial of order $i$ and $P_j(\cos{\theta})$ are Legendre Polynomials. Because our solution is symmetric about a reflection of the equatorial plane $\theta = \pi/2$, we need only consider even Legendre polynomials. We then fit these models to our numerical solutions to determine the constants $(a_{i,j},b_{i,j},c_{i,j},d_{i,j},e_{i,j})$ on the grid domain $x \in [0,1]$ and $\theta \in [0, \pi/2]$. The fitting order of our models is determined by systematically increasing the polynomial order of each function until the residual between the numerical solution and the model saturates. The best-fit coefficients $(a_{i,j}, b_{k,l}, c_{m,n}, d_{p,q}, e_{r,s})$ are available in a {\texttt{Mathematica}} file at \url{https://github.com/sullivanandrew/XPDES}. 

\begin{figure*}[p]
\begin{center}
\resizebox{8cm}{!}{\includegraphics{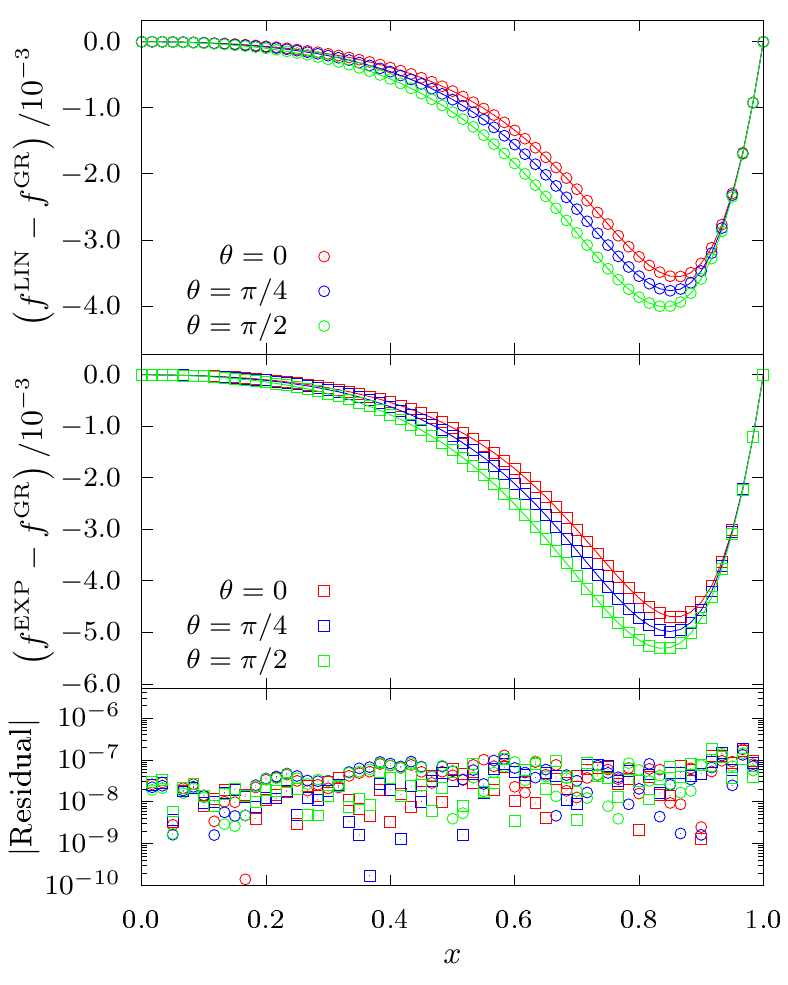}}
\resizebox{8cm}{!}{\includegraphics{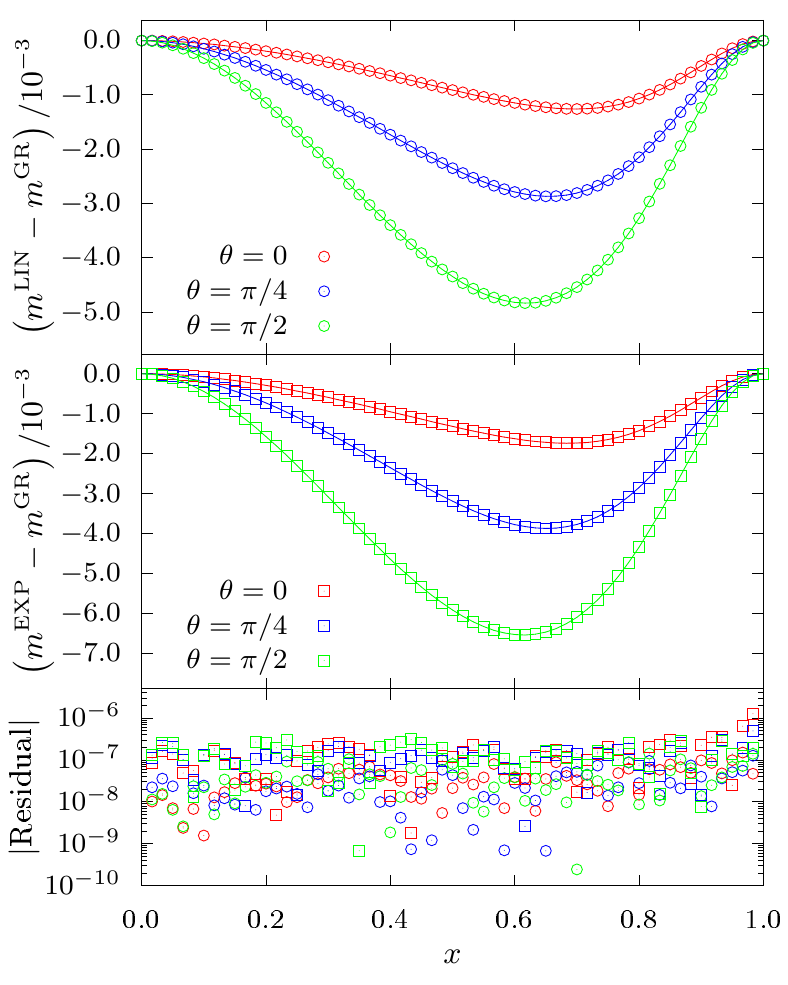}}
\\
\resizebox{8cm}{!}{\includegraphics{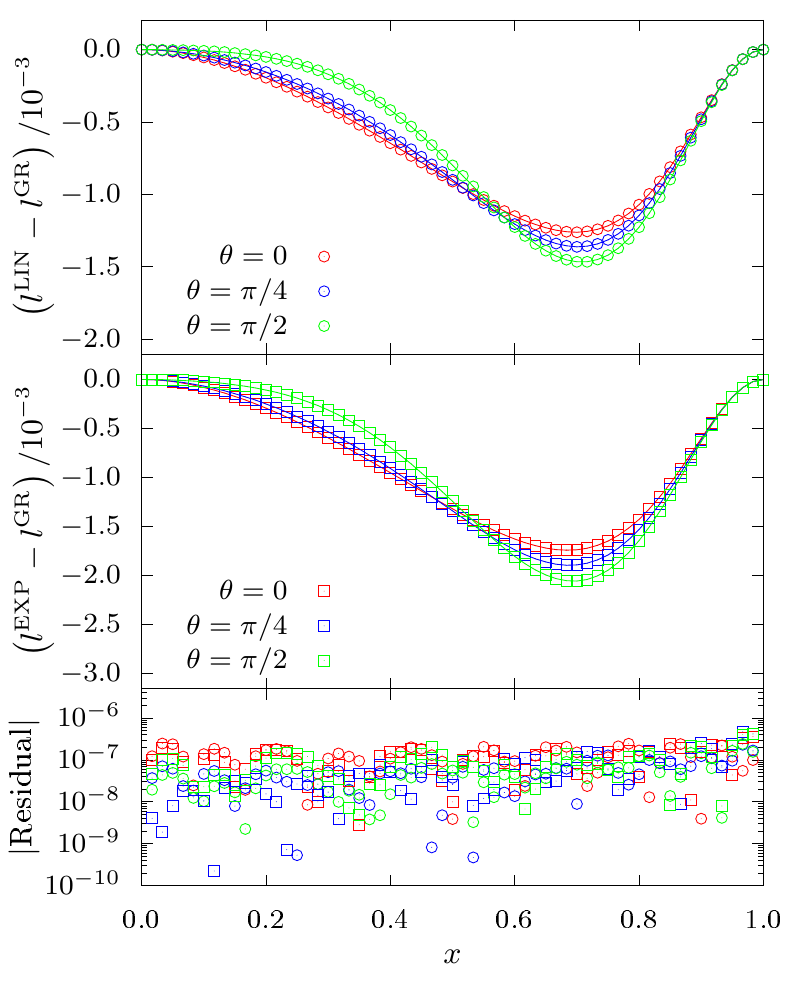}}
\resizebox{8cm}{!}{\includegraphics{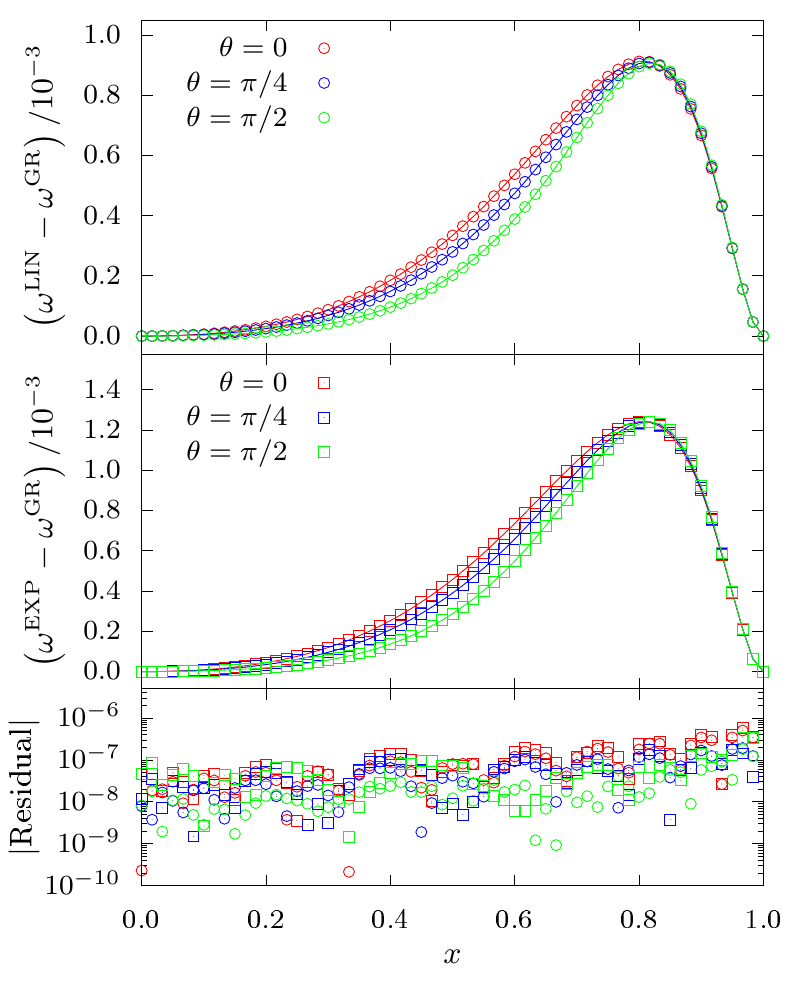}}
\caption{\label{fig:StackFitg} 
(Color online)
Difference between metric elements and the Kerr metric using the linear coupling function (top half of each panel) and exponential coupling function (middle of each panel) for three selected angles. Overlaid is the analytical fit denoted by the solid line while the dots denote the numerical solution for a black hole of $\bar{\alpha} = 0.5$ and $\Omega_{\Hz} = 0.0667$. The bottom of each panel is the absolute value of the residual between the analytical fit and the numerical data. Here we show the metric components $f$ (top left panel), $m$ (top right panel), $l$ (bottom left panel) and $\omega$ (bottom right panel) for three angles $\theta = 0,\pi/4, \pi/2$ denoted by the red, blue, and green colors respectively. Notice the residual remains below the specified tolerance on the numerical solution on the entire domain.}
\end{center}
\end{figure*}

\begin{figure}[h]
\begin{center}
\resizebox{8cm}{!}{\includegraphics{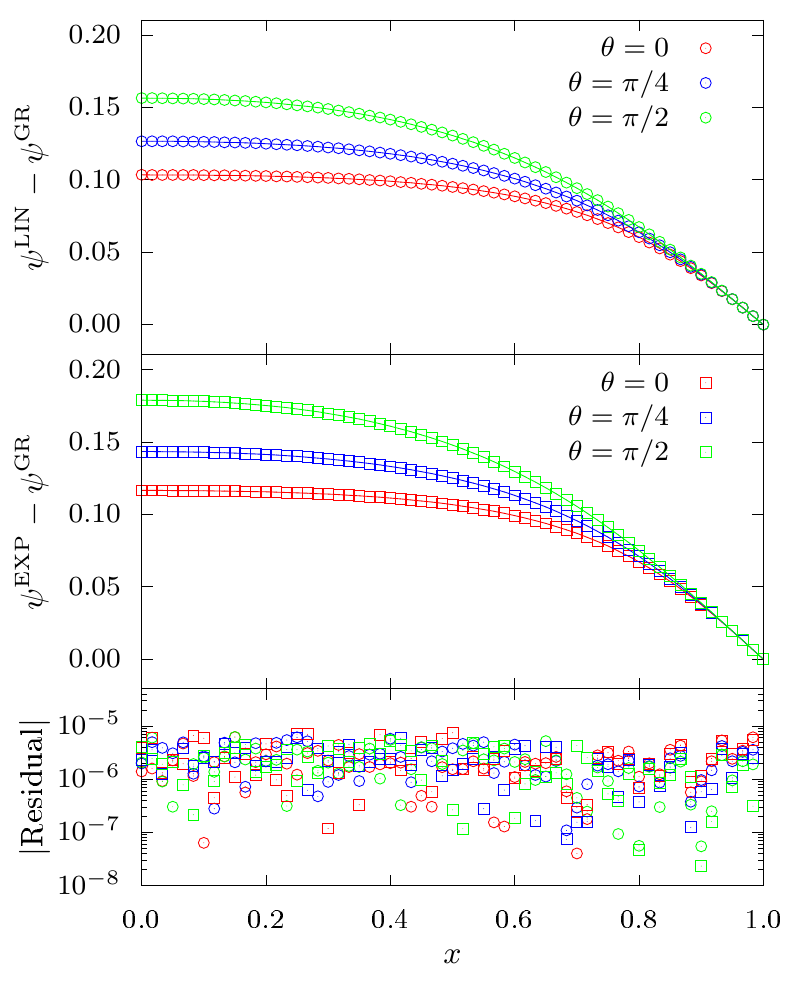}}
\caption{\label{fig:StackFitp} 
(Color online)
The scalar field using the linear coupling function and its analytical fit for $\bar{\alpha} = 0.5$ and $\Omega_{\Hz} = 0.0667$ and the absolute residual between the analytical fit and the numerical data (bottom). The analytic fit for three angles $\theta = 0,\pi/4, \pi/2$ are denoted by the red, blue, and green solid lines respectively while the dots denote the numerical solution.}
\end{center}
\end{figure}

We plot the difference between both the numerical solutions and the fitted models for a coupling of $\bar{\alpha} = 0.5$ and horizon angular velocity $\Omega_{\Hz} = 0.0667$ 
 to the Kerr solution as well as the residuals between the models and the numerical data for the metric components and the scalar field for the linear sGB and EdGB solutions in Fig.~\ref{fig:StackFitg} and~\ref{fig:StackFitp}. We find that the residual between the models and the numerical data is always below the specified tolerance on the numerical solution of $\mcl{O}(10^{-5})$. Thus the fitted models can be treated as ``exact" for practical applications to the specified tolerance.


%
%

\subsection{Marginal Stable Circular Orbits}

To numerically calculate the location of the marginal stable circular orbits (MSCO) around a stationary, axially symmetric black hole, we begin with a generic metric ansatz of the form
\be
ds^2 = \tn{g}{_t_t} dt^2 + \tn{g}{_r_r} dr^2 + \tn{g}{_\theta_\theta} d\theta^2 + \tn{g}{_\phi_\phi} d\phi^2 + 2 \tn{g}{_t_\phi} dt d\phi.
\ee
The two killing vectors of our spacetime $\tn{t}{^\mu}$ and $\tn{\phi}{^\mu}$ correspond to the reduced energy $E$ and angular momentum $L$ of the particle,
\bal
\begin{split}
\label{eq:EandL}
E &= -\tn{t}{_\mu} \frac{d \tn{x}{^\mu}}{d \lambda} = - \tn{g}{_t_t} \dot{t} - \tn{g}{_t_\phi} \dot{\phi}, \\
L &= \tn{\phi}{_\mu} \frac{d \tn{x}{^\mu}}{d \lambda} = \tn{g}{_t_\phi} \dot{t} + \tn{g}{_\phi_\phi} \dot{\phi},
\end{split}
\end{align}
which can be combined to obtain expressions for $\dot{t}$ and $\dot{\phi}$,
\bal
\begin{split}
\label{eq:dott-dotp}
\dot{t} &= \frac{E \tn{g}{_t_t} + L \tn{g}{_t_\phi}}{\tn{g}{_t_\phi}^2 - \tn{g}{_t_t} \tn{g}{_\phi_\phi}},\\
\dot{\phi} &= - \frac{E \tn{g}{_t_\phi} + L \tn{g}{_t_t}}{\tn{g}{_t_\phi}^2 - \tn{g}{_t_t} \tn{g}{_\phi_\phi}}.
\end{split}
\end{align}

If we consider orbits constrained to the equatorial plane $\theta=\pi/2$, the four-velocity normalization condition becomes
\be
-\epsilon = \tn{g}{_t_t} \dot{t}^2 + \tn{g}{_r_r} \dot{r}^2 + \tn{g}{_\phi_\phi} \dot{\phi}^2 + 2 \tn{g}{_t_\phi} \dot{t} \dot{\phi},
\ee
where $\epsilon = 0$ for photon and $\epsilon = 1$ for massive particles. Inserting $\dot{t}$ and $\dot{\phi}$, we can solve for $\dot{r}^2$ and define an effective potential $U_{\eff}$ given by
\be
\dot{r}^2 = \frac{1}{\tn{g}{_r_r}} \lb -\epsilon + \frac{E^2 \tn{g}{_\phi_\phi} + 2 E L \tn{g}{_t_\phi} + L^2 \tn{g}{_t_t}}{\tn{g}{_t_\phi}^2 - \tn{g}{_t_t} \tn{g}{_\phi_\phi}} \rb \equiv U_{\eff}.
\ee
The condition for a circular orbit is $\dot{r} = 0 = \ddot{r}$, and by differentiating
\be
\frac{d}{d \lambda} \dot{r}^2 = 2 \dot{r} \ddot{r} = \frac{d U_{\eff}}{d r} \dot{r} \rarr \ddot{r} = \frac{d U_{\eff}}{d r} = 0,
\ee
we find that these two conditions imply that the effective potential and its derivative must vanish. 

These two conditions can be rearranged into two algebraic equations that must be simultaneously satisfied:
\bal
\label{eq:Ueff}
E^2 \tn{g}{_\phi_\phi} + 2 E L \tn{g}{_t_\phi} + L^2 \tn{g}{_t_t} -\epsilon \lb \tn{g}{_t_\phi}^2 - \tn{g}{_t_t} \tn{g}{_\phi_\phi} \rb &= 0, \\
\label{eq:dUeff}
E^2 \tn{g}{_\phi_\phi}^{\prime} + 2 E L \tn{g}{_t_\phi}^{\prime} + L^2 \tn{g}{_t_t}^{\prime} -\epsilon \lb \tn{g}{_t_\phi}^2 - \tn{g}{_t_t} \tn{g}{_\phi_\phi} \rb^{\prime} &= 0,
\end{align}
where the primes denote radial derivatives e.g. $\tn{g}{_t_t}^{\prime} = {d \tn{g}{_t_t}}/{d r}$, evaluated at the radius $r$. We now turn to specific cases of these marginal stable circular orbits: the light ring, and the innermost stable circular orbit.

\subsection{Light Ring}

For a photon, $\epsilon = 0$ and Eq.~\eqref{eq:Ueff} can be solved quadratically for $E$ or $L$
\be
L = E \lb \frac{\tn{g}{_t_\phi} \pm \sqrt{\tn{g}{_t_\phi}^2 - \tn{g}{_t_t} \tn{g}{_\phi_\phi}}}{\tn{g}{_t_t}} \rb,
\ee
and this result can be inserted into Eq.~\eqref{eq:dUeff} to obtain the equation
\bal
\begin{split}
\label{eq:LReq}
\tn{g}{_\phi_\phi}^{\prime} &+ 2 \tn{g}{_t_\phi}^{\prime} \lb \frac{\tn{g}{_t_\phi} \pm \sqrt{\tn{g}{_t_\phi}^2 - \tn{g}{_t_t} \tn{g}{_\phi_\phi}}}{\tn{g}{_t_t}} \rb \\
&+ \tn{g}{_t_t}^{\prime} \lb \frac{\tn{g}{_t_\phi} \pm \sqrt{\tn{g}{_t_\phi}^2 - \tn{g}{_t_t} \tn{g}{_\phi_\phi}}}{\tn{g}{_t_t}} \rb^2 = 0,
\end{split}
\end{align}
which is to be evaluated at a radius $r$; the smallest root of the above equation is the location of the light ring. 

Once we insert the metric functions known analytically or numerically, we only need to determine the root of the above equation to find the location of the light ring. 
With our nonlinear numerical solutions, we can approximate the derivatives using our Newton interpolation polynomial and use a Newton-Raphson algorithm to find the root. The results were presented in the top of Fig.~\ref{fig:ISCOLR}. We find that the increased scalar charge of a black hole in sGB due to an increasing in coupling $\bar{\alpha}$ will push the location of the ISCO away from the horizon ($\delta_{\mrm{ISCO}} > 0$), but increasing the rotation of the black hole pushes the ISCO towards the horizon ($\delta_{\mrm{ISCO}} < 0$) as in GR. These competing effects can even cause the fractional shift in the ISCO to vanish in the special case that they exactly cancel. We also find that the magnitude of the fractional change in the location of the ISCO is suppressed by increasing the angular momentum. 

\subsection{Innermost Stable Circular Orbit}

For a massive particle, $\epsilon = 1$ and the innermost stable circular orbit is located at the saddle point of the effective potential, specifically when $U_{\eff}^{\prime \prime} = 0$. This adds another condition that must be satisfied and another equation analogous to Eq.~\eqref{eq:dUeff}, namely
\be
\label{eq:d2Ueff}
E^2 \tn{g}{_\phi_\phi}^{\prime \prime} + 2 E L \tn{g}{_t_\phi}^{\prime \prime} + L^2 \tn{g}{_t_t}^{\prime \prime} - \epsilon \lb \tn{g}{_t_\phi}^2 - \tn{g}{_t_t} \tn{g}{_\phi_\phi} \rb^{\prime \prime} = 0.
\ee

To find the ISCO, we begin by solving Eq.~\eqref{eq:Ueff} quadratically for $L$ similar to the approach taken for the light ring, namely
\be
\label{eq:LofE}
L = \frac{E \tn{g}{_t_\phi} \pm \sqrt{\lb E^2 + \epsilon \tn{g}{_t_t} \rb  \lb \tn{g}{_t_\phi}^2 - \tn{g}{_t_t} \tn{g}{_\phi_\phi} \rb}}{\tn{g}{_t_t}}.
\ee
We insert this expression into Eq.~\eqref{eq:dUeff} and solve for $E$ as a function of only the metric and its first derivatives
\be
\label{eq:Emetric}
E = E \lsb \tn{g}{_\mu_\nu}, \tn{g}{_\mu_\nu}^{\prime} \rsb,
\ee
which can also be substituted back into Eq.~\eqref{eq:LofE} to obtain $L$ also as a function of only the metric and its first derivatives
\be
L = L \lsb \tn{g}{_\mu_\nu}, \tn{g}{_\mu_\nu}^{\prime} \rsb.
\ee
These expressions are calculated in the symbolic manipulation software \textit{Maple 2018} available at \url{https://github.com/sullivanandrew/XPDES} and will not be presented here. Finally, we can substitute both of these into Eq.~\eqref{eq:d2Ueff} to obtain a second order equation, the smallest root of which is the ISCO. 

As in the case with the light ring, this is done numerically with a Newton-Raphson algorithm and the result is shown on the bottom plot of Fig.~\ref{fig:ISCOLR}. We find a similar competing effect between the scalar charge and the angular momentum as with the ISCO. The light ring appears to be more sensitive to the effect from the angular momentum due to its closer proximity to the event horizon than the ISCO. 

\section{Conclusions}
\label{sec:concs}

We have presented here a numerical infrastructure to calculate the exterior spacetimes of rotating black holes in a wide class of modified theories of gravity. We have validated this infrastructure by obtaining the Kerr solutions in GR and by direct comparison with a rotating, weak-coupling perturbative numerical solution in sGB gravity. We then compared the full nonlinear solutions to rotating black holes to find the deviations from GR in the metric functions and the physical observables such as the mass and angular momentum. 

We have also used these numerical solutions to construct analytical fitted models that reproduce the data to within the accuracy of the solutions, and calculated other physical observables like location of the the ISCO and light ring. We have found that the solutions in linear sGB gravity are very closely approximated by the perturbative weak-coupling expansion and that these solutions differ quite drastically from the corresponding solutions in EdGB gravity. We have also found that the deviations of rotating black holes from the Kerr spacetime become increasingly suppressed for  larger black hole spins, as the deviations sourced by the scalar charge begin to become dominated by the gravitational effects of the angular momentum.

The analytical fitted models constructed from these solutions can be used to calculate other astrophysical observables such as accretion disks around black holes~\cite{Abramowicz2013} or black hole shadows~\cite{0264-9381-35-23-235002}. These solutions can also be used as a background to study polar and axial perturbations to predict the quasinormal mode spectrum of scalar Gauss-Bonnet black holes~\cite{Blazquez-Salcedo:2016enn,PhysRevD.99.104077,PhysRevD.100.044061,10.1007/978-3-319-10488-1_19}. These can then be compared to gravitational wave ringdown observations of merging black holes of future detectors to place constraints on a variety of modified gravity theories.

\acknowledgments

We are grateful for the computational support of the Hyalite High-Performance Computing System, operated and supported by University Information Technology Research Cyberinfrastructure at Montana State University. A.~S.~ and N.~Y.~would like to acknowledge support from NSF PHY-1759615 and NASA grant 80NSSC18K1352. T.~P.~S.~acknowledges partial support from the STFC Consolidated Grant No.~ST/P000703/1 and  networking support from the COST Action GWverse CA16104.

\bibliographystyle{apsrev}
\bibliography{NumerAXIBiblio}
\end{document}